\documentclass[aps,showpacs,eqsecnum,10pt]{revtex4}

\usepackage{graphicx}
\usepackage{amsmath}
\usepackage{amssymb}

\begin{document}


\title{Transport properties in the $d$-density-wave state
in an external magnetic field: The Wiedemann-Franz law}

\author{S.G.~Sharapov$^{1}$}
\email{Sergei.Sharapov@unine.ch}
\author{V.P.~Gusynin$^{1,2}$}
\email{vgusynin@bitp.kiev.ua}

\author{H.~Beck$^1$}%
\email{Hans.Beck@unine.ch}
\homepage{http://www.unine.ch/phys/theocond/}

\affiliation{
        $^1$Institut de Physique,
        Universit\'e de Neuch\^atel, 2000 Neuch\^atel, Switzerland\\
        $^2$Bogolyubov Institute for Theoretical Physics,
        Metrologicheskaya Str. 14-b, Kiev, 03143, Ukraine}

\date{\today }

\begin{abstract}
We derive the electrical and thermal conductivities of the
$d$-density-wave state in an external magnetic field $B$ in the
low-temperature regime and in the presence of impurities. We show
that in the zero-temperature limit, $T \to 0$, the Wiedemann-Franz
(WF) law remains intact irrespectively of the value of the applied
field and chemical potential $\mu$ as soon as there is scattering
from impurities. For finite $T \lesssim |\mu|$ the WF law
violation is possible and it is enhanced by the external field.
Depending on the relative values of $B$ and $T$ the electrical
conduction may dominate the heat conduction and vice versa. For
$\mu \gg T$ the WF is restored even in the presence of the
external field.
\end{abstract}




\maketitle

\section{Introduction}

A recent experiment of Hill {\em et al.}  \cite{Hill:2001:Nature}
that measured electrical and thermal conductivities of the
optimally doped copper oxide superconductor
Pr$_{2-x}$Ce$_x$CuO$_4$ (PCCO) in its normal state found striking
deviations from the Wiedemann-Franz (WF) law. The electron-doped
PCCO compound with $T_c \simeq 20 \mbox{K}$ near the optimal $x
=0.15$ doping is the analog of the hole doped material
La$_{2-x}$Sr$_x$CuO$_{4-y}$ (LSCO), which is also optimally doped
at approximately $x=0.15$ with a maximum $T_{c} \simeq 40
\mbox{K}$. Lower $T_c$ in PCCO makes it particularly convenient
from a technical point of view since an external magnetic field is
necessary to destroy superconductivity to study its low
temperature transport properties in the normal state. While for
LSCO the upper critical field is $\simeq 50 \mbox{T}$, in the
optimally doped PCCO the superconductivity is already destroyed at
$8 \mbox{T}$ \cite{Hill:2001:Nature}. The choice of overdoped
compounds that can be driven into the normal state by a relatively
small magnetic field $\sim 13 \mbox{T}$ is wider and the hole
overdoped system Tl$_2$Ba$_2$CuO$_{6+\delta}$ (Tl-2201) with $T_c
= 15 \mbox{K}$ was studied very recently by Proust {\em et al.}
\cite{Proust:2002:PRL}. They verified that in the overdoped
Tl-2201 WF law holds perfectly.

The WF law is one of the basic properties of a Fermi liquid, reflecting
the fact that the ability of a quasiparticle to transport heat
is the same as its ability to transport charge, provided it cannot
lose energy through collisions. The WF law states that the ratio
of the heat conductivity $\kappa$ to the electrical conductivity
$\sigma$ of a metal is  a universal constant:
\begin{equation}
L_0 \equiv \frac{\kappa}{\sigma T} = \frac{\pi^2}{3}
\left( \frac{k_B}{e}\right)^2,
\end{equation}
where $k_B$ is the Boltzmann constant, $e$ is electron charge, and
$L_0 = 2.45 \times 10^{-8} \mbox{W} \Omega \mbox{K}^{-2}$ is
Sommerfeld's value for the Lorenz ratio $L \equiv  \kappa/(\sigma
T)$. To be more precise, one should also specify the temperature
range where the WF law holds. Strictly speaking this law is proven
only in the limit $T \to 0$ and for a small concentration of
impurities \cite{Langer:1962:PR}. In a less strict sense it is
often implied that the WF law is valid for $k_B T \ll \mu$, where
$\mu$ is chemical potential. In this case one can provide some
qualitative arguments \cite{Taylor.book} that if the scattering
from impurities does not strongly depend on the energy the WF will
still be valid. Moreover, for  $k_B T \ll \mu$ and elastic
electron scattering these arguments can also be extended to the
case of a strong ($\hbar \omega_L \gg k_B T$, $\hbar \omega_L \gg
\Gamma$) quantizing magnetic field \cite{Smrcka:1977:JPC}, where
$\hbar \omega_L$ is the distance between Landau levels and
$\Gamma$ is the impurity scattering rate. Thus, if these
conditions are fulfilled, any violation of the WF law would
indicate that there is a breakdown of the picture based on the
Fermi liquid theory.

One of the possible theoretical interpretations of the WF law
breakdown is that the quasiparticle fractionalizes into separate
spin and charge. This separation can be investigated using various
models and approaches, so here we mention only the most recent
studies done in the context of the WF law violation in cuprates.
Specifically, the WF law violation has been studied for a
large-$N$ limit of the $t-J$ model on the square lattice
\cite{Houghton:2002:PRB}. Another examination of the WF law was
done by Yang and Nayak (YN) \cite{Yang:2002:PRB} and also by Kim
and Carbotte (KC) \cite{Kim:2002:PRB} within the phenomenological
$d$-density-wave (DDW) picture.

\begin{figure}[h]
\centering{
\includegraphics[width=8cm]{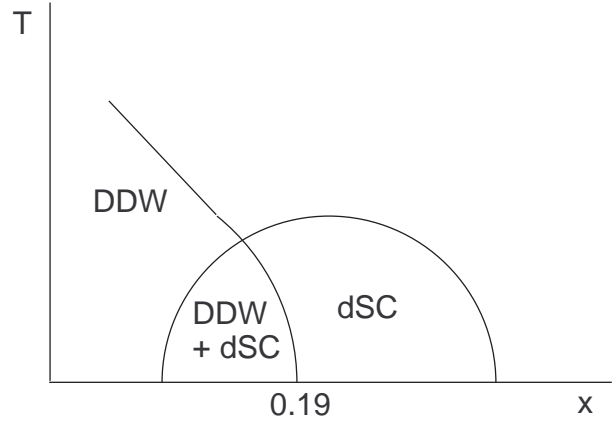}}
\caption{Schematic phase diagram of cuprates within the DDW scenario.}
\label{fig:1}
\end{figure}

The DDW scenario proposed in Ref.~\cite{Chakravarty:2001:PRB} is
based on the assumption that the pseudogap phenomenon
\cite{Timusk:1999:RPP} in high-$T_c$ cuprates is the result of the
development of another order parameter called the DDW order
\cite{Nayak:2000:PRB} that has $d$-wave symmetry:
\begin{equation}
\label{DDW.def}
\langle c_{s}^{\dagger} (t,\mathbf{k} + \mathbf{Q})
c_{s^{\prime}}(t,\mathbf{k}) \rangle = i \Phi_{\mathbf{Q}} f(\mathbf{k})
\delta_{s s^{\prime}}
\end{equation}
where $s, s^{\prime} = \pm$ is the spin index, $f(\mathbf{k}) =
\cos k_x a - \cos k_y a $, $\mathbf{Q} = (\pi/a, \pi/ a)$, and $a$
is the lattice constant. In comparison to the $d$-wave
superconducting order the DDW order parameter appears to be rather
exotic because  it breaks the time reversal, translation by one
lattice spacing and rotation by $\pi/2$  symmetries, but respects
any combination of two of these. The underlying reason for this
breaking is that there are countercirculating currents in the DDW
ground state. The schematic phase diagram for the DDW pseudogap
scenario is shown in Fig.~\ref{fig:1}. For a hole doping larger
than the critical value $x_{c} \approx 0.19$
\cite{Tallon:2001:PC}, the DDW order is presumed to disappear,
whereas for the underdoped system DDW order coexists with $d$-wave
superconductivity ($d$SC).

One of the unusual features of the DDW state is that for a
half-filled system the chemical potential of the nodal
quasiparticles participating in the electrical and thermal
transport can be small or even zero, i.e., $|\mu| < k_B T$, which
violates the usual conditions of the WF law validity. Indeed,
exactly in the limit $\mu =0$ the WF law is strongly violated in
the extremely clean limit \cite{Yang:2002:PRB}. There is no WF law
violation in the $T =0$ limit for finite $\mu$ and/or $\Gamma$
\cite{Yang:2002:PRB,Kim:2002:PRB}. For finite temperatures the WF
violation depends on the impurity scattering \cite{Kim:2002:PRB}:
in the Born limit (for a constant impurity scattering rate) there
is no change in the WF law, but in the unitary limit for the
frequency dependent scattering rate the WF law is violated, but
only for $|\mu|$ smaller that the DDW gap. When $\mu$ is increased
sufficiently, the Lorenz number becomes approximately equal to its
conventional value and its temperature dependence is small.

While in general the validity of the DDW pseudogap scenario is
still questionable, it is important to scrutinize all its
theoretical consequences. One of the opportunities  is to study
possible WF law violations using the DDW model, so that these
results can be compared with the experimental results of
Refs.~\cite{Hill:2001:Nature,Proust:2002:PRL}. Although this
investigation to a large extent had already been done by YN
\cite{Yang:2002:PRB} and KC \cite{Kim:2002:PRB}, both these papers
do not consider {\em the presence of the external magnetic
field\/} which is an essential ingredient of the experiments
\cite{Hill:2001:Nature,Proust:2002:PRL}. Indeed, in these
experiments
 the DDW state appears in the magnetic field
in the underdoped regime at low $T$  when the
superconductivity is destroyed.

In this paper we study the WF law for the DDW model {\em
including\/} the external magnetic field for a constant impurity
scattering rate, paying special attention to the regime $|\mu|
\lesssim k_B T$ where the violation of the WF law is expected.
Although there are some formal similarities between the DDW and
$d$-wave superconducting states, the present problem is much
simpler because there are no vortices and no superflows
surrounding them in the DDW state. The $U(1)$ symmetry in the DDW
state also remains unbroken, so that the electromagnetic field
enters into the effective low-energy theory in the same way as in
QED$_{2+1}$. Thus we avoid the famous problem of {\em Dirac
Landau-level mixing} (see
Refs.~\cite{Vafek:2001:PRB,Mel'nikov:2002:book}).

We begin by presenting in Sec.~\ref{sec:model} the model DDW
Hamiltonian and rewriting it using two-component spinors. Since
there is a discrepancy between the expressions for the electrical
current used by YN \cite{Yang:2002:PRB} and KC
\cite{Kim:2002:PRB}, we also derive the electrical current
operator and show that in the nodal approximation it reduces to
the current of Ref.~ \cite{Yang:2002:PRB}. Then we obtain the
low-energy effective action (Lagrangian) for the nodal
quasiparticles in the presence of an external electromagnetic
field, consider its discrete symmetries, and rewrite the
Lagrangian in the Dirac form. In Sec.~\ref{sec:Green.magnetic} we
consider the Green's function of nodal quasiparticles in an
applied field (the calculational details are given in
Appendix~\ref{sec:A}).  In Appendix~\ref{sec:B}, using this
Green's function, we derive a generalized polarization bubble that
can be applied to the calculation of electrical and thermal
conductivities. In Secs.~\ref{sec:electrical} and
\ref{sec:thermal} we obtain general expressions for electrical and
thermal conductivities in the external field. These general
expressions are further analyzed in detail by analytical and
numerical methods. In Sec.~\ref{sec:WF} we investigate the
implications for the WF law that follow from the results presented
in the previous sections. Conclusions are discussed in
Sec.~\ref{sec:conclusions}, where we give a concise summary of the
obtained results and compare them with experiment. Some useful
integrals are given in Appendix~\ref{sec:C}.

\section{\lowercase{$d$}-density wave Hamiltonian and its effective
low energy form}
\label{sec:model}

We start with the mean-field Hamiltonian for the DDW state
\cite{Nayak:2000:PRB,Yang:2002:PRB,Nersesyan:1989:JLTP}
\begin{equation}
\label{Hamiltonian.DDW}
H^{\mathrm{DDW}} = \int_{\mathrm{BZ}} \frac{d^2 k}{(2 \pi)^2}
[(\varepsilon(\mathbf{k}) - \mu)
c_{s}^{\dagger}(\mathbf{k}) c_{s}(\mathbf{k})
+ i D(\mathbf{k})
c_{s}^{\dagger}(\mathbf{k}) c_{s}(\mathbf{k} + \mathbf{Q}) ],
\end{equation}
where the single particle energy is $\varepsilon({\mathbf{k}}) = -
2 t (\cos k_x a + \cos k_y a) - 4 t^{\prime} \cos k_x a \cos k_y
a$ with $t$, $t^{\prime}$ being the hopping parameters, $\mu$ is
the chemical potential, $D(\mathbf{k}) = \frac{D_0}{2}(\cos k_x a
- \cos k_y a )$ is the $d$-density-wave gap and $\mathbf{Q} =
(\pi/a, \pi/ a)$ is the wave vector at which the density-wave
ordering takes place. The integral is over the full Brillouin
zone. Throughout the paper $\hbar = k_{B} = c= 1$ units are
chosen, unless stated explicitly otherwise.

Due to the presence of the factor $i$ in the second term of the
Hamiltonian (\ref{Hamiltonian.DDW}), it breaks time-reversal symmetry,
where
the time-reversal operation is defined as \cite{Enz.book}
\begin{equation}
\label{time.reversal.Enz}
c_{s}(\mathbf{k}) \to
\mathcal{T} c_{s}(\mathbf{k}) \mathcal{T}^{-1} = i s c_{-s}(-\mathbf{k}),
\qquad
c_{s}^{\dagger}(\mathbf{k}) \to
\mathcal{T} c_{s}^{\dagger}(\mathbf{k}) \mathcal{T}^{-1} =
- i s c_{-s}^{\dagger}(-\mathbf{k})
\end{equation}
and the time-reversal operator $\mathcal{T}$ is unitary and
antilinear. Nevertheless the Hamiltonian (\ref{Hamiltonian.DDW})
preserves a combined symmetry of the time reversal and a
translation by one lattice spacing $\tilde{\mathcal{T}} =
\mathcal{T} T_{\mathbf{a}}$:
\begin{equation}
\label{time-translation}
c_s(\mathbf{k}) \to
\tilde{\mathcal{T}} c_s(\mathbf{k}) \tilde{\mathcal{T}}^{-1} =
i e^{-i \mathbf{k} \mathbf{a}} c_{-s}(-\mathbf{k}).
\end{equation}
Since the order parameter also breaks  translational symmetry by
one lattice spacing, it is convenient to halve the Brillouin zone
as shown in Fig.~\ref{fig:2} and form a two-component electron
operator
\begin{equation}
\label{Nambu.variables}
\chi_s(t, \mathbf{k}) = \left( \begin{array}{c}
c_{s}(t, \mathbf{k}) \\
c_{s}(t, \mathbf{k} + \mathbf{Q})
\end{array} \right), \qquad
\chi_s^{\dagger}(t, \mathbf{k}) = \left( \begin{array}{cc}
c_{s}^{\dagger}(t, \mathbf{k})
\quad c_{s}^{\dagger}(t, \mathbf{k + Q})
\end{array} \right).
\end{equation}
Then the mean-field Hamiltonian (\ref{Hamiltonian.DDW}) in terms of $\chi$ becomes
\begin{equation}
\label{Hamiltonian.DDW.spinor}
H^{\mathrm{DDW}}  = \int_{\mathrm{RBZ}} \frac{d^2 k}{(2 \pi)^2}
\chi_{s}^{\dagger}(t, \mathbf{k}) \left[
\frac{1}{2}[\varepsilon(\mathbf{k}) + \varepsilon(\mathbf{k} + \mathbf{Q})] - \mu
+ \frac{1}{2}[\varepsilon(\mathbf{k}) - \varepsilon(\mathbf{k} + \mathbf{Q})]
\sigma_3 - D(\mathbf{k}) \sigma_2
\right]  \chi_{s}(t, \mathbf{k}),
\end{equation}
where $\sigma_i$ are Pauli matrices and the integral is over the
reduced Brillouin zone (RBZ). The symmetry operation
(\ref{time-translation}), for example, is
\begin{equation}
\label{time-reversal.spinor}
\chi_s(t, \mathbf{k}) \to
\tilde{\mathcal{T}} \chi_s(t, \mathbf{k}) \tilde{\mathcal{T}}^{-1}
= -  e^{-i \mathbf{k} \mathbf{a}}
\sigma_2 \chi_{-s}(-t, -\mathbf{k} - \mathbf{Q}).
\end{equation}
In what follows we will consider the more simple case $t^{\prime} =0$,
so that after employing the nesting property
$\varepsilon(\mathbf{k} + \mathbf{Q}) = -
\varepsilon(\mathbf{k})$ the Hamiltonian (\ref{Hamiltonian.DDW.spinor})
takes a simple form
\begin{equation}
\label{Hamiltonian.DDW.matrix}
H^{\mathrm{DDW}}  = \int_{\mathrm{RBZ}} \frac{d^2 k}{(2 \pi)^2}
\chi_{s}^{\dagger}(t, \mathbf{k}) \left[ H_0(\mathbf{k}) - \mu
\right] \chi_{s}(t, \mathbf{k}),
\end{equation}
where
\begin{equation}
\label{H}
H_0(\mathbf{k}) =\varepsilon(\mathbf{k}) \sigma_3 - D(\mathbf{k}) \sigma_2.
\end{equation}

\subsection{Electrical and energy current operators}

As we already mentioned the expressions for the electrical current
used in Refs.~\cite{Yang:2002:PRB} and \cite{Kim:2002:PRB} are
different. Although quantitatively this difference for
high-temperature superconductors is expected to be rather small
[see after Eq.~(\ref{Green.nodal})], the extra term in the current
present in Ref.~\cite{Yang:2002:PRB} is of qualitative importance,
so we need to discuss its origin.

To find out electrical current operator $\mathbf{j}$ for
Hamiltonian (\ref{Hamiltonian.DDW}) we apply the method similar to
that of Refs.~\cite{Durst:2000:PRB} and \cite{Kim:2002:PRB}, where
it was used to determine the heat current in the $d$-wave
superconducting and DDW states, respectively. The idea of the
derivation is to make the expression for the current operator
compatible with the charge conservation law,
\begin{equation}
\label{conserve}
\frac{\partial \rho(t, \mathbf{r})}{\partial t} +
\nabla \cdot \mathbf{j}(\mathbf{t, \mathbf{r}}) =0,
\end{equation}
and with the equations of motion for field operators,
\begin{equation}
\label{motion}
\begin{split}
 i \frac{\partial c_{s}^{\dagger}(\mathbf{k})}{\partial t} =
 [ c_{s}^{\dagger}(\mathbf{k}), H] =
 -c_{s}^{\dagger}(\mathbf{k})
[\varepsilon(\mathbf{k}) - \mu]
 -  c_{s}^{\dagger}(\mathbf{k}  - \mathbf{Q})
i D(\mathbf{k}  - \mathbf{Q}).
\end{split}
\end{equation}
Writing Eq.~(\ref{conserve}) in the momentum space, one obtains
\begin{equation}
\label{conserve.momentum}
\frac{\partial \rho(t,\mathbf{\mathbf{q}})}{\partial t} +
i \mathbf{q} \cdot \mathbf{j}(t, \mathbf{q}) =0,
\end{equation}
where the Fourier transform of
the charge density is
\begin{equation}
\rho(t, \mathbf{q}) = e \int_{\mathrm{BZ}} \frac{d^2 k}{(2 \pi)^2}
c_{s}^{\dagger}(\mathbf{k}) c_{s}(\mathbf{k}  + \mathbf{q}).
\end{equation}
For its derivative we obtain
\begin{equation}
\begin{split}
\frac{\partial \rho(t, \mathbf{q})}{\partial t} =
 -i e \int_{\mathrm{BZ}} \frac{d^2 k}{(2 \pi)^2}  \left[
c_{s}^{\dagger}(\mathbf{k})
c_{s}(\mathbf{k}  + \mathbf{q}) \mathbf{q}
\frac{\partial \varepsilon(\mathbf{k})}{\partial \mathbf{k}}
+ c_{s}^{\dagger}(\mathbf{k})
c_{s}(\mathbf{k}  + \mathbf{q}+ \mathbf{Q}) i \mathbf{q}
\frac{\partial D(\mathbf{k})}{\partial \mathbf{k}} \right],
\qquad \mathbf{q} \to 0,
\end{split}
\end{equation}
so that the electrical current is
\begin{equation}
\label{electric.current.DDW}
\begin{split}
\mathbf{j} (t, \mathbf{0}) & =
e \int_{\mathrm{BZ}} \frac{d^2 k}{(2 \pi)^2}  \left[
c_{s}^{\dagger}(\mathbf{k}) c_{s}(\mathbf{k})
\frac{\partial \varepsilon(\mathbf{k})}{\partial \mathbf{k}}
+ c_{s}^{\dagger}(\mathbf{k} )
c_{s}(\mathbf{k} + \mathbf{Q}) i
\frac{\partial D(\mathbf{k})}{\partial \mathbf{k}} \right] \\
& = e \int_{\mathrm{RBZ}} \frac{d^2 k}{(2 \pi)^2}
\chi_s^{\dagger}(\mathbf{k}) \mathbf{V}(\mathbf{k})
 \chi_s(\mathbf{k}),
\end{split}
\end{equation}
where we introduced the generalized velocity operator
that depends of the Fermi, $\mathbf{v}_F$, and gap, $\mathbf{v}_D$, velocities
\begin{equation}
\label{velocity.general}
\mathbf{V}(\mathbf{k}) =
\mathbf{v}_F(\mathbf{k}) \sigma_3  +
\mathbf{v}_D(\mathbf{k}) \sigma_2,
\qquad
\mathbf{v}_F(\mathbf{k}) \equiv
\frac{\partial \epsilon(\mathbf{k})}{\partial \mathbf{k}}, \quad
\mathbf{v}_D(\mathbf{k}) \equiv -\frac{\partial D(\mathbf{k})}{\partial \mathbf{k}}.
\end{equation}
In the last equality in Eq.~(\ref{electric.current.DDW})
we used the two-component form and relied
on the nesting property $\varepsilon(\mathbf{k} + \mathbf{Q}) = -
\varepsilon(\mathbf{k})$ and $D(\mathbf{k} + \mathbf{Q}) = - D(\mathbf{k})$.

This form of the electrical current operator reduces to the
current used by YN \cite{Yang:2002:PRB} after the nodal
approximation is made. The last term of
Eq.~(\ref{electric.current.DDW}), $\sim \mathbf{v}_D$, is due to
the fact that the DDW gap in the mean-field Hamiltonian
(\ref{Hamiltonian.DDW}) is $\mathbf{k}$-dependent. The expression
for the electrical current used by KC contains only one term $\sim
\mathbf{v}_F$. In this case, calculating electrical conductivity
in the bare bubble approximation is not consistent with
charge-current conservation, since such a bare vertex does not
satisfy the Ward identity when quasiparticle self-energy has a
nontrivial momentum dependence. Instead one should use a vertex
that complies with the Ward identity.

The derivation of the energy current operator from the Hamiltonian
(\ref{Hamiltonian.DDW})
is quite similar to the derivation of the electrical current operator
(\ref{electric.current.DDW}), so here we only outline the main steps.
This current is calculated from the corresponding
continuity equation for the energy density  and the equations of
motion (\ref{motion}) (we set there $\mu =0$ because the energy density is considered)
that give
\begin{equation}
\label{heat.current}
\mathbf{j}^E (t, \mathbf{0}) = \int_{\mathrm{BZ}} \frac{d^2 k}{(2 \pi)^2}
\left[\varepsilon(\mathbf{k})
\mathbf{v}_F(\mathbf{k}) +
D(\mathbf{k})  \mathbf{v}_D(\mathbf{k}) \right]
c_s^{\dagger}(\mathbf{k}) c_s(\mathbf{k}),
\end{equation}
or in the spinor form
\begin{equation}
\mathbf{j}^E (t, \mathbf{0}) = \frac{1}{2}
\int_{\mathrm{RBZ}} \frac{d^2 k}{(2 \pi)^2}
\chi^{\dagger}(t, \mathbf{k}) \{H_0(\mathbf{k}), \mathbf{V}(\mathbf{k}) \}
\chi(t, \mathbf{k}).
\end{equation}
Then using again the equations of motion we finally get
\begin{equation}
\label{heat.current2.B=0}
\mathbf{j}^E (\Omega, \mathbf{0})  =
\int_{\mathrm{RBZ}} \frac{d^2 k}{(2 \pi)^2} \int d \omega \,
\left(\omega + \frac{\Omega}{2} \right)
\chi^{\dagger}(\omega, \mathbf{k}) \mathbf{V}(\mathbf{k}) \chi(\omega+ \Omega, \mathbf{k}).
\end{equation}
One can easily see that Eq.~(\ref{heat.current2.B=0}) agrees with
the corresponding energy current obtained in Ref.~\cite{Kim:2002:PRB}.

\subsection{Effective low-energy nodal action}

The action $S$ corresponding to the Hamiltonian (\ref{Hamiltonian.DDW.matrix}) is
\begin{equation}
\begin{split}
S & =  - \int \limits_{0}^{\beta} d \tau \left[ \int d^2 r
\chi_{s}^{\dagger} (\tau,\mathbf{r}) \partial_{\tau}
\chi_{s} (\tau,\mathbf{r}) + H^{\mathrm{DDW}}(\tau)
\right] \\
&  = - \int \limits_{0}^{\beta} d \tau \int_{\mathrm{RBZ}} \frac{d^2 k}{(2 \pi)^2}
\chi_{s}^{\dagger} (\tau,\mathbf{k})
\left[ \partial_{\tau} - \mu + \varepsilon(\mathbf{k}) \sigma_3 -
D(\mathbf{k}) \sigma_2 \right] \chi_{s} (\tau,\mathbf{k}),
\end{split}
\end{equation}
where $\tau$ is the imaginary time.
Hence the Green's function of the DDW state reads
\begin{equation}
\label{Green.common}
G(i\omega, \mathbf{k}) =  \frac{(i\omega + \mu)\hat{I} +
\varepsilon(\mathbf{k}) \sigma_3 -  D(\mathbf{k}) \sigma_2}
{(i \omega + \mu)^2 - E^{2}(\mathbf{k})}, \qquad
E(\mathbf{k}) =  \sqrt{\varepsilon^2(\mathbf{k}) + D^2(\mathbf{k})},
\end{equation}
where $i \omega = i  (2n+1) \pi T $ is fermionic (odd) Matsubara frequency.

\begin{figure}[h]
\centering{
\includegraphics[width=8cm]{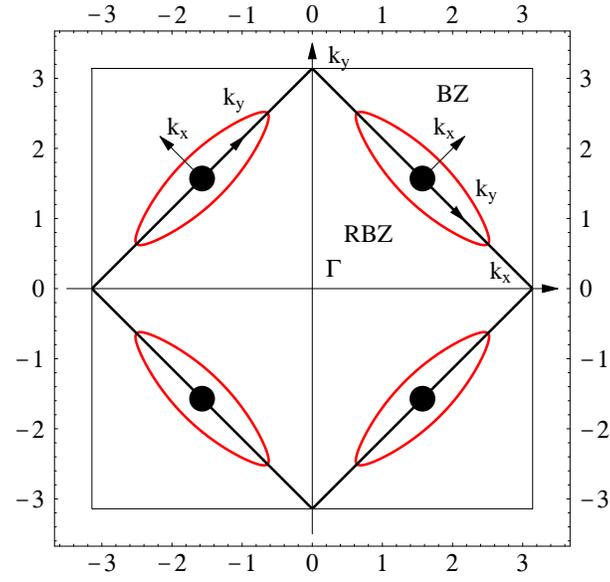}}
\caption{Fermi surface in the DDW state. The hole pockets are centered
around the points $(\pm \pi/2, \pm \pi/2)$. We choose the local nodal
coordinate systems as shown.}
\label{fig:2}
\end{figure}

Linearizing the spectrum about the four nodes $\mathbf{N} = (\pm
\pi/2a, \pm \pi/2a )$ at half-filling ($\mu =0$), one obtains the
Green's function (GF) for a given node (we choose the local nodal
coordinate systems as shown in Fig.~\ref{fig:2})
\begin{equation}
\label{Green.nodal}
G_{\mathrm{node}}(i\omega, \mathbf{k}) =  \frac{(i\omega + \mu)\hat{I} +
v_F k_x \sigma_3 +  v_D k_y \sigma_2}
{(i \omega + \mu)^2 - E^2(\mathbf{k})}, \qquad
E(\mathbf{k}) =  \sqrt{v_F^2 k_x^2 + v_D^2 k_y^2},
\end{equation}
where the Fermi velocity for half-filling $v_F = |\partial
\varepsilon(\mathbf{k})/\partial \mathbf{k} |_{\mathbf{k} =
\mathbf{N}} | = 2 \sqrt{2} t a$ and the DDW gap velocity $v_{D} =
|\partial D(\mathbf{k})/\partial \mathbf{k} |_{\mathbf{k} =
\mathbf{N}} | = \frac{1}{\sqrt{2}} D_0 a$. It is important to
stress that this way of linearization is different from the nodal
approximation used to describe $d$SC \cite{Durst:2000:PRB} when
the expansion is done around the nodal (``Dirac'') points defined
as an intersection of the Fermi surface $\varepsilon (\mathbf{k})
- \mu =0$ and the nodal lines $D(\mathbf{k}) = 0$. For the DDW
case one always expands around half-filling $\varepsilon
(\mathbf{k}) =0$ and $D(\mathbf{k}) = 0$, so that for nonzero
$\mu$ the nodes transform into small pockets (see
Fig.~\ref{fig:2}) which are cross sections of the ``Dirac cone''.
Thus only at exactly half-filling is there a correspondence
between DDW and $d$SC cases. There are the following estimates
\cite{Morr:2002:PRL} of the model parameters, $t = 300$ meV and
$D_0 = 50$meV, which give the ratio $v_{F}/v_{D} = 24$. This shows
that the second term of the electrical current
(\ref{electric.current.DDW}), which is proportional to $v_{D}$, is
indeed small. The value of the chemical potential corresponding to
a hole doping of 10\% is \cite{Morr:2002:PRL} $\mu = 0.91 t$.

The linearized Dirac action that corresponds to the Green's function (\ref{Green.nodal})
can be written as
\begin{equation}
\label{S.nodal}
\begin{split}
S_{\mathrm{node}} = - \int \limits_{0}^{\beta} d \tau \int d^2 r
\chi_{s}^{\dagger}(\tau, \mathbf{r})
[\hat{I} (\partial_{\tau} - \mu) - v_F \sigma_3 i \partial_x -
v_D \sigma_2 i \partial_y ] \chi_{s}(\tau, \mathbf{r}) .
\end{split}
\end{equation}
Dealing with Eqs.~(\ref{Green.nodal}) and (\ref{S.nodal}), one
should not forget that all physical quantities involve the
summation over the 4 nodes present in the original Green's
function (\ref{Green.common}). In what follows we will use the
nodal action (\ref{S.nodal}) instead of the original DDW
Hamiltonian (\ref{Hamiltonian.DDW}). This approximation is well
justified for $T \ll D_0$. As we will see below, this
approximation and the fact that the electromagnetic field enters
the theory through the minimal coupling prescription will allow us
to apply the results of QED$_{2+1}$.

\subsection{Discrete symmetries of the nodal Lagrangian}

Since the action (\ref{S.nodal}) does not directly correspond to
the common QED$_{2+1}$ form used in the literature
\cite{Dittrich.book} (see also Ref.~\cite{Jackiw:1981:PRD}), we
present below the transformations for the discrete symmetries. Let
us consider  only one node and suppress the spin index $s$, so
that in real time $t = - i \tau$ the action $S$ reads
\begin{equation}
S_{\mathrm{node}} = i \int d t \int d^2 r \mathcal{L},
\end{equation}
with
\begin{equation}
\label{Dirac.Lagrangian.real}
\mathcal{L} =  \chi^{\dagger}(t, \mathbf{r})
\left[i \hat{I} (\partial_{t} - i e A_0(t, \mathbf{r})) + \mu
+ i v_F \sigma_3 \left(\partial_x - i \frac{e}{c}A_1(t, \mathbf{r})\right)
+ i v_D \sigma_2 \left( \partial_y - i\frac{e}{c}A_2(t, \mathbf{r})\right)
\right] \chi (t, \mathbf{r}),
\end{equation}
where $A_{\nu}(t, \mathbf{r})$, $\nu =0,1,2$ is the
electromagnetic field vector potential. This is a {\em highly
nontrivial fact\/} in that the vector potential can be inserted in
Eq.~(\ref{Dirac.Lagrangian.real}) using the {\em minimal coupling
prescription\/} of QED. Only due to this fact can one  apply the
results of  QED$_{2+1}$ for the description of the DDW state.
Moreover, this is only correct for the DDW state
\cite{Yang:2002:PRB} but not for the $d$SC state, where the last
term of the Lagrangian (\ref{Dirac.Lagrangian.real}) does not
couple with the electromagnetic field. It has to be admitted that
the way of inserting the electromagnetic field is not just related
to the nodal approximation and can be traced back to the
expression (\ref{electric.current.DDW}) for the electrical current
that can only be obtained when the vector potential enters both
$\varepsilon(\mathbf{k})$ and $D(\mathbf{k})$ terms in
Eq.~(\ref{Hamiltonian.DDW}). Finally we note that the symmetries
of the Lagrangian have to be analyzed in real time where one can
distinguish time and spatial coordinates.

\subsubsection{Parity}

In $2+1$ dimensions, parity corresponds to inverting one axis
(since the inversion of both axes would be rotation by $\pi$):
$P \mathbf{r} = \mathbf{r}^{\prime}$, where
$\mathbf{r} = (x,y)$ and $\mathbf{r}^{\prime} = (x, -y)$.
The corresponding operation on the two-component spinor and on
the gauge field is
\begin{equation}
\begin{split}
&  \chi(\tau, \mathbf{r}) \to
\mathcal{P} \chi(\tau, \mathbf{r}) \mathcal{P}^{-1} =
\sigma_3 \chi(\tau, \mathrm{r}^{\prime}),
\qquad
\chi^{\dagger}(t, \mathbf{r}) \to
\mathcal{P} \chi^{\dagger}(t, \mathbf{r}) \mathcal{P}^{-1} =
\chi^{\dagger}(t, \mathrm{r}^{\prime}) \sigma_3, \\
& A_0(t, \mathbf{r}) \to
\mathcal{P} A_{0}(t, \mathbf{r}) \mathcal{P}^{-1} = A_{0}(t, \mathbf{r}^{\prime}),
\quad
A_1(t, \mathbf{r}) \to
\mathcal{P} A_{1}(t, \mathbf{r}) \mathcal{P}^{-1} = A_{1}(t, \mathbf{r}^{\prime}),
\quad
A_2(t, \mathbf{r}) \to
\mathcal{P} A_{2}(t, \mathbf{r}) \mathcal{P}^{-1} = -A_{2}(t, \mathbf{r}^{\prime}).
\end{split}
\end{equation}
One can verify that the Lagrangian (\ref{Dirac.Lagrangian.real})
is invariant under this parity transformation while possible mass
terms with $\sigma_1$ and $\sigma_2$ break parity:
\begin{equation}
\begin{split}
& \chi^{\dagger}(t, \mathbf{r}) \sigma_1 \chi(t, \mathbf{r}) \to
\mathcal{P}\chi^{\dagger}(t, \mathbf{r}) \sigma_1 \chi(t, \mathbf{r}) \mathcal{P}^{-1}
=  - \chi^{\dagger}(t, \mathbf{r}^{\prime})
\sigma_1 \chi(t, \mathbf{r}^{\prime}); \\
&  \chi^{\dagger}(t, \mathbf{r}) \sigma_2 \chi(t, \mathbf{r}) \to
\mathcal{P}\chi^{\dagger}(t, \mathbf{r}) \sigma_2 \chi(t, \mathbf{r}) \mathcal{P}^{-1}
=  - \chi^{\dagger}(t, \mathbf{r}^{\prime})
\sigma_2 \chi(t, \mathbf{r}^{\prime}).
\end{split}
\end{equation}

\subsubsection{Charge conjugation}

Charge conjugation,
\begin{equation}
\begin{split}
&  \chi (t, \mathbf{r}) \to \mathcal{C} \chi(t, \mathbf{r})
\mathcal{C}^{-1} = \sigma_3 [\chi^{\dagger}(t,\mathbf{r})]^T,
\qquad \chi^{\dagger} (t, \mathbf{r}) \to \mathcal{C}
\chi^{\dagger} (t, \mathbf{r}) \mathcal{C}^{-1} =
[\chi(t,\mathbf{r})]^T \sigma_3, \\
& A_\nu(t, \mathbf{r}) \to
\mathcal{C} A_{\nu}(t, \mathbf{r}) \mathcal{C}^{-1} = -A_{\nu}(t, \mathbf{r}),
\qquad \nu =0,1,2
\end{split}
\end{equation}
leaves the equations of motion invariant. It is easy to check
that the  Lagrangian (\ref{Dirac.Lagrangian.real})
is invariant under this transformation only at half-filling ($\mu = 0$)
while away from half-filling the term with the chemical potential breaks $C$
\begin{equation}
\mu \chi^{\dagger}(t, \mathbf{r}) \hat{I}\chi(t, \mathbf{r}) \to
\mu \mathcal{C}\chi^{\dagger}(t, \mathbf{r}) \hat{I}\chi(t, \mathbf{r}) \mathcal{C}^{-1}
= - \mu  \chi^{\dagger}(t, \mathbf{r}) \hat{I} \chi(t, \mathbf{r}).
\end{equation}
One can also check that the mass term with $\sigma_2$ also breaks
$C$, whereas $\sigma_1$ mass term does not break $C$.

\subsubsection{Time-reversal symmetry}

The operation of time reversal for the electromagnetic field is
defined as
\begin{equation}
A_0(t, \mathbf{r}) \to \mathcal{T} A_{0}(t, \mathbf{r}) \mathcal{T}^{-1} =
A_{0}(-t, \mathbf{r}), \qquad
A_{1,2}(t, \mathbf{r}) \to \mathcal{T} A_{1,2}(t, \mathbf{r}) \mathcal{T}^{-1} =
-A_{1,2}(-t, \mathbf{r}).
\end{equation}
We determine a matrix $T$
responsible for the time-reversal transformation of spinors
\begin{equation}
\label{time.reversal.Dirac}
\chi (t, \mathbf{r}) \to
\mathcal{T} \chi(t, \mathbf{r}) \mathcal{T}^{-1} =
T \chi(-t,\mathbf{r}), \qquad
\chi^{\dagger} (t, \mathbf{r})  \to
\mathcal{T} \chi^{\dagger} (t, \mathbf{r}) \mathcal{T}^{-1} =
\chi^{\dagger}(-t,\mathbf{r}) T
\end{equation}
demanding that the current components transform as follows
\begin{equation}
\chi^{\dagger}(t, \mathbf{r}) \alpha_{\nu} \chi(t, \mathbf{r}) \to
\chi^{T \dagger}(-t, \mathbf{r}) \alpha_{\nu}^{\ast} \chi^{T}(-t, \mathbf{r}) =
\chi^{\dagger}(-t, \mathbf{r}) T \alpha_{\nu}^{\ast} T \chi(-t, \mathbf{r}) =
\chi^{\dagger}(-t, \mathbf{r}) \tilde{\alpha}_{\nu} \chi(-t, \mathbf{r}),
\end{equation}
where
\begin{equation}
\label{alpha.nu}
\alpha_{\nu} = (\hat{I}, \sigma_3, \sigma_2),
\qquad \tilde{\alpha}_{\nu} = (\alpha_0, - \alpha_1,  -\alpha_2).
\end{equation}

The time-reversal operation (\ref{time.reversal.Dirac})
can be written down using $T = \sigma_2$ and one can see that
the Lagrangian $\mathcal{L}$ is invariant with respect to this
symmetry
\begin{equation}
\mathcal{L}[\chi^{\dagger}(t, \mathbf{r}), \chi(t, \mathbf{r}), A_{\nu}(t, \mathbf{r})]
\to
\mathcal{L}[\chi^{T \dagger}(-t, \mathbf{r}), \chi^{T}(-t, \mathbf{r}),
A_{\nu}^{T}(-t, \mathbf{r})]
=  \mathcal{L}[\chi^{\dagger}(t, \mathbf{r}),  \chi(t, \mathbf{r}),
A_{\nu}(t, \mathbf{r})],
\end{equation}
while the $\sigma_1$  and $\sigma_2$ mass terms would break it
\begin{equation}
\chi^{\dagger}(t, \mathbf{r}) \sigma_{1,2} \chi(t, \mathbf{r}) \to
\mathcal{T}\chi^{\dagger}(t, \mathbf{r}) \sigma_{1,2} \chi(t, \mathbf{r}) \mathcal{T}^{-1}
= -  \chi^{\dagger}(-t, \mathbf{r})  \sigma_{1,2} \chi(-t, \mathbf{r}).
\end{equation}
One can check that if we use notations corresponding to those of
Ref.~\cite{Jackiw:1981:PRD}, $P$ and $T$ operations introduced
here agree with the transformations given in
Ref.~\cite{Jackiw:1981:PRD}. The time-reversal symmetry operation
(\ref{time.reversal.Dirac}) for the continuum nodal Lagrangian
should be identified with the combined symmetry operation
(\ref{time-reversal.spinor}) that leaves the Hamiltonian
(\ref{Hamiltonian.DDW.spinor}) invariant. We note that, in
principle, the time-reversal transformation should also flip the
spin $s \to - s$, but since all considered expressions assume
summation over two spin components, we do not include this spin
flip in the transformation of Dirac spinors.

Parity and time-reversal symmetries can still be broken by the mass (gap) term
$\sim \sigma_1$. In particular, this mass term may be generated in an
external magnetic field due to the so called {\em magnetic catalysis\/} phenomenon
\cite{Gusynin:1995:PRD}.

\subsection{Final form of the nodal Lagrangian}

We choose the vector potential for the external magnetic field $\mathbf{B}$ in the
symmetric Poincar\'e gauge $\mathbf{A} = (-\frac{B}{2}x_2 , \frac{B}{2} x_1 )$,
so that the field $\mathbf{B}$ is perpendicular to CuO$_2$ planes.
To consider also a possibility of magnetic catalysis \cite{Gusynin:1995:PRD}
we add to the Lagrangian
(\ref{Dirac.Lagrangian.real}) the interaction term
originating  from particle-hole attraction
\begin{equation}
\label{L.int}
\mathcal{L}_{\mathrm{int}} =  \frac{g}{2} (\chi^{\dagger}(x) \sigma_1 \chi(x))^2,
\qquad x = (t, \mathbf{r}).
\end{equation}
The simplest way to treat the interaction (\ref{L.int})
is to introduce Hubbard-Stratonovich field
$\varphi(x) = - g \chi^{\dagger}(x) \sigma_1 \chi(x)$, so that
the Lagrangian (\ref{Dirac.Lagrangian.real}) becomes
\begin{equation}
\label{Lagrangian.alpha}
\mathcal{L} =  \chi^{\dagger}(x) [i \alpha_{\nu} D_{\nu} -\sigma_1 \varphi(x)  ]\chi (x)-
\frac{\varphi^2(x)}{2g},
\end{equation}
where the covariant derivatives  are
\begin{equation}
\label{long.derivative}
D_{\nu} =
\begin{cases}
\hbar \partial_t  - i e A_{0}(x), &      \nu =0,\\
v_{F} \left(\hbar \partial_x - i \frac{e}{c} A_{1}(x) \right), &  \nu =1, \\
v_{\Delta} \left( \hbar \partial_y - i \frac{e}{c} A_{2}(x)\right), &  \nu =2,
\end{cases}
\end{equation}
and  $\alpha$ matrices were defined in
Eq.~(\ref{alpha.nu}).

Finally, to simplify further calculations and to make a direct link with
QED$_{2+1}$ it is convenient to introduce Dirac conjugated spinor
$\bar{\chi}(x) = \chi^{\dagger}(x) \sigma_1$ and rewrite the
Lagrangian (\ref{Lagrangian.alpha}) as
\begin{equation}
\label{Dirac.conjugated.Lagrangian}
\mathcal{L} =  \bar{\chi}(x) [i \gamma^{\nu} D_{\nu} - \varphi(x)  ]\chi (x)-
\frac{\varphi^2(x)}{2g},
\end{equation}
where the $\gamma$ matrices are
\begin{equation}
\gamma^{\nu} = (\sigma_1,  - i \sigma_2, i \sigma_3),
\qquad \{\gamma^\mu, \gamma^{\nu}\} = 2 \hat{I} g^{\mu \nu}, \qquad
g^{\mu \nu} = \mbox{diag}(1,-1,-1).
\end{equation}
Note that the Zeeman term, if necessary, can be added explicitly
both to the original Hamiltonian (\ref{Hamiltonian.DDW}) and the Lagrangian
(\ref{Dirac.conjugated.Lagrangian}). Here, however, this term is neglected.

\section{Green's function in an external magnetic field}
\label{sec:Green.magnetic}

To derive the transport coefficients we need an explicit
representation for the fermionic Green's function in an external
field. The calculation, following the Schwinger (proper time)
approach \cite{Schwinger:1951:PR}, is sketched in
Appendix~\ref{sec:A} and the result is
(we set for convenience $v_F = v_D =1$
and restore them when it is needed according to the prescription given in
Appendix~\ref{sec:B})
\begin{equation}
\label{Schwinger.representation}
S(x-y) = \exp \left( i e \int \limits_y^x A_{\lambda}^{\mathrm{ext}} d z^{\lambda} \right)
{\tilde S}(x-y),
\end{equation}
where the translation invariant part ${\tilde S}$ in the Matsubara
frequency-momentum representation is given by
\begin{equation}
\label{Schwinger.representation.translation}
\begin{split}
{\tilde S}(i \omega, \mathbf{p}) = &  -
\int \limits_{0}^{\infty} d s \exp \left[-s \left(\Delta^2 -(i \omega)^2 +
\mathbf{p}^2 \frac{\tanh (eBs)}{eBs}\right) \right] \\
& \times
\left[ \left(i \omega \gamma^0 - p_1 \gamma^1 - p_2 \gamma^2 + \Delta +
\frac{1}{i}(p_2 \gamma^1 - p_1 \gamma^2) \tanh (e B s) \right)
\left(1 + \frac{1}{i} \gamma^1 \gamma^2 \tanh (e B s) \right)  \right].
\end{split}
\end{equation}
Here we replaced the Hubbard-Stratonovich field $\varphi(x)$ by
its mean-field value $\Delta = \langle \varphi (x)\rangle$. The
chemical potential $\mu$ has to be taken into account via the
shift $i\omega  \to i \omega + \mu$ [see
Eq.~(\ref{G.translation.invariant}) in Appendix~\ref{sec:B}].

As already mentioned, the gap
$\Delta$ can be generated by an external magnetic field \cite{Gusynin:1995:PRD}.
The value of the gap has to be determined from the minimum condition
of the corresponding effective potential, see e.g.
\cite{Gusynin:1995:PRD,Ferrer:2002,Gorbar:2002:PRB}.
Here, however,  we are mainly interested in  the case
of $\Delta = 0$ and  will concentrate on $\mu$ and $T$
dependences of the Lorenz ratio $L$.

The propagator (\ref{Schwinger.representation.translation}) can be
decomposed over the Landau level poles
\cite{Gusynin:1995:PRD,Chodos:1990:PRD} (see also
Appendix~\ref{sec:A} for the details):
\begin{equation}
\label{Landau.levels}
{\tilde S}(i \omega, \mathbf{p}) =
\exp\left( - \frac{\mathbf{p}^2}{|e B|}\right) \sum_{n=0}^{\infty} (-1)^n
\frac{S_n(B,i \omega, \mathbf{p})}{(i\omega)^2 - \Delta^2 - 2 |e B| n},
\end{equation}
where
\begin{equation}
\label{Sn}
S_n(B,i \omega, \mathbf{p}) =
  2 (\Delta + i \omega \gamma^0) \left[ P_{-} L_n \left(2\frac{\mathbf{p}^2}{|eB|}\right)
- P_{+} L_{n-1}\left(2 \frac{\mathbf{p}^2}{|eB|} \right)
\right]
 + 4 (p_1 \gamma^1 + p_2 \gamma^2) L_{n-1}^1 \left(2 \frac{\mathbf{p}^2}{|eB|} \right)
\end{equation}
with $P_{\pm} = [1\pm  \mbox{sgn} (e B)i \gamma^1 \gamma^2]/2 = (1
\mp \sigma_1)/2$ being projectors and $L_n$, $L_n^1$ Laguerre's
polynomials ($L_{-1}^1 \equiv 0$). In what follows for convenience
we take $e B > 0$. The spectral function $A_D(\omega, \mathbf{p})$
associated with the translation-invariant part ${\tilde S}$ of the
Green's function is defined as
\begin{equation}
\label{spectral.def}
A_D(\omega, \mathbf{p}) = \frac{1}{2 \pi i} \left[
{\tilde S}^A(\omega - i0, \mathbf{p}) - {\tilde S}^R(\omega + i0, \mathbf{p}) \right],
\end{equation}
where the retarded, ${\tilde S}^R$, and advanced, ${\tilde S}^A$, Greens' functions are
${\tilde S}^R(\omega + i 0, \mathbf{p}) =
{\tilde S}(i \omega \to \omega + i 0, \mathbf{p})$
and
${\tilde S}^A(\omega - i 0, \mathbf{p}) =
{\tilde S}(i \omega \to \omega - i 0, \mathbf{p})$.
Using the definition (\ref{spectral.def}) one obtains
the spectral function associated with the ``Dirac'' fermion
Green's function $\langle \chi \bar{\chi} \rangle$  (\ref{Landau.levels})
for a clean system
\begin{equation}
\label{AD.clean}
A_D(\omega, \mathbf{p}) =
\exp\left( - \frac{\mathbf{p}^2}{e B}\right) \sum_{n=0}^{\infty} (-1)^n
\left[ \frac{(\gamma^0 M_n + \Delta) f_1(\mathbf{p}) +  f_2(\mathbf{p})}{2 M_n}
\delta(\omega - M_n) +
\frac{(\gamma^0 M_n - \Delta) f_1(\mathbf{p}) -  f_2(\mathbf{p})}{2 M_n}
\delta(\omega + M_n)
\right],
\end{equation}
where $M_n = \sqrt{\Delta^2 + 2 e B n}$
and
\begin{equation}
\label{f}
f_1(\mathbf{p}) = 2\left[
P_{-} L_n \left(2\frac{\mathbf{p}^2}{eB}\right)
- P_{+} L_{n-1}\left(2 \frac{\mathbf{p}^2}{eB} \right) \right], \qquad
f_2(\mathbf{p}) = 4 (p_1 \gamma^1 + p_2 \gamma^2)
L^{1}_{n-1}\left(2 \frac{\mathbf{p}^2}{eB} \right).
\end{equation}

To consider transport phenomena for a more realistic case one
should introduce into the theory the effect of scattering on
impurities. In general this can be done by considering dressed
fermion propagators that include the  self-energy $\Sigma(\omega)$
due to the scattering on impurities. This self-energy in  turn has
to be found self-consistently by solving the corresponding
Schwinger-Dyson equation. This scheme was in fact used in the
Ref.~\cite{Kim:2002:PRB}, and the WF violation obtained by KC at
finite temperatures in the unitary limit is essentially due to the
nontrivial frequency dependence of the scattering rate
$\Gamma(\omega) = -\mbox{Im} \Sigma^R(\omega)$. As mentioned in
Introduction, here we choose the case of constant width $\Gamma =
\Gamma(\omega = 0) = 1/(2 \tau)$, $\tau$ being a mean free time of
quasiparticles, so that the $\delta$-like quasiparticle peaks  in
Eq.~(\ref{AD.clean}) acquire a Lorentzian shape:
\begin{equation}
\label{Gamma}
\delta(\omega \pm M_n) \to \frac{\Gamma}{\pi} \frac{1}{(\omega \pm M_n)^2 + \Gamma^2}.
\end{equation}
Such a broadening of Landau levels was found to be a rather good approximation
valid in not very strong magnetic fields \cite{Prange.book}.

Finally,  we establish a link between the  Green's function
(\ref{Schwinger.representation}) considered here and the Green's
function (\ref{Green.common}) used in the preceding section.
Taking into account that the average $\langle \chi \chi^{\dagger}
\rangle$  is related to the ``Dirac's average'' via $\langle \chi
\chi^{\dagger} \rangle = \langle \chi \bar{\chi} \rangle
\gamma^{0}$ one obtains for $B = 0$:
\begin{equation}
\label{G.B=0}
G(i\omega, \mathbf{p}) = {\tilde S}(i \omega_m,\mathbf{p}; B =0) \gamma^0 =
\frac{i \omega + p_1 \sigma_3 + p_2 \sigma_2 + \Delta \sigma_1}
{(i \omega)^2 - \mathbf{p}^2 - \Delta^2},
\end{equation}
so that restoring $v_F$, $v_D$, replacing $i\omega_n \to i
\omega_n + \mu$, and setting $\Delta =0$ we recover the Green's
function (\ref{Green.nodal}).

\section{Electrical conductivity}
\label{sec:electrical}

\subsection{General expression for electrical conductivity}

The frequency dependent longitudinal electrical conductivity can
be calculated from the current-current correlation function
\begin{equation}
\label{el.cur-cur.tensor}
\Pi^{CC}_{\alpha \beta}(i \Omega) = - \int \limits_{0}^{\beta} d \tau e^{i \Omega \tau}
\langle T_{\tau} j_{\alpha}^{\dagger}(\tau, \mathbf{0})
j_{\beta}(0, \mathbf{0})
\rangle
\end{equation}
by means of the Kubo formula \cite{Mahan:book}
\begin{equation}
\label{el.cur-cur.scalar}
\sigma(\Omega) = -  \frac{\mbox{Im} \Pi_{R}^{CC}(\Omega +i 0)}{\Omega},
\end{equation}
where $\Pi_{R}^{CC}(\Omega + i0) = \Pi^{CC}(i \Omega \to \Omega +
i0)$ is the longitudinal polarization [see
Eq.~(\ref{tensor2scalar})].

The expression for the electrical current operator for the
Hamiltonian (\ref{Hamiltonian.DDW}) was already derived in
Eq.~(\ref{electric.current.DDW}). Having the current, we are in a
position to calculate the current-current correlation function
(\ref{el.cur-cur.tensor}), which is given by the bubble diagram.
The influence of the impurity vertex corrections on the transport
properties of another ($d$SC) nodal system was considered in
Ref.~\cite{Durst:2000:PRB}. The calculation of the bare bubble
polarization function is in fact similar for all transport
coefficients in $d$SC and DDW nodal liquids, so that it is
convenient to define a generalized polarization function  $\Pi^{g
g^{\prime}}(\Omega)$ that depends on the coupling parameters $g$,
$g^{\prime}$ and the generalized velocity
$\mathbf{V}(\mathbf{k})$. The polarization function is calculated
in Appendix~\ref{sec:B}. Using the general result
(\ref{bubble.final.tensor}) for the case of interest, $g =
g^{\prime} = e$ in the limit $\Omega \to 0$, we find that the dc
conductivity of isotropic system is given by
\begin{equation}
\label{sigma.like.Langer}
\sigma =
\pi e^2 \int_{\mathrm{RBZ}} \frac{d^2 k}{(2 \pi)^2}
\int \limits_{-\infty}^{\infty} d\omega
(-n_F^{\prime}(\omega - \mu))
\mbox{tr} \left[ A(\omega, {\bf k}) V_{\alpha}(\mathbf{k})
A(\omega, {\bf k}) V_{\alpha}(\mathbf{k})  \right] ,
\end{equation}
where $-n_{F}^{\prime}(\omega -\mu) = (1/4T) \cosh^{-2}[(\omega -\mu)/2T]$
is the derivative of the Fermi distribution and summation over the dummy
variable is implied.

At this point it is instructive to compare
Eq.~(\ref{sigma.like.Langer}) with the expression for the
conductivity obtained by Langer [Eq.~(4.8) of
Ref.~\cite{Langer:1962a:PR}] in the bubble approximation. Langer's
expression has the same form as Eq.~(\ref{sigma.like.Langer})
except the matrix-valued $e V_{\alpha}(\mathbf{k})$ is replaced by
the vertex with coinciding fermion momenta and energies
\begin{equation}
\label{Lambda.Langer}
\Lambda_i(\mathbf{k}, \omega) = e \frac{k_i}{m} - e
\frac{\partial}{\partial k_i} \Sigma(\mathbf{k}, \omega),
\end{equation}
where $\Sigma(\mathbf{k}, \omega)$ is the usual self-energy [he
considered the quadratic dispersion law, so that $\partial
\epsilon(\mathbf{k})/\partial k_i = k_i/m$]. In fact
Eq.~(\ref{Lambda.Langer}) is a direct consequence of the Ward
identity. Thus knowing the self-energy one immediately obtains the
dc conductivity in the bubble approximation. Furthermore, the
bubble conductivity of Ref.~\cite{Langer:1962a:PR} becomes exact
in the zero-temperature limit.

Since in the mean-field approximation $\Sigma(\mathbf{k}, \omega)$
is replaced by the DDW gap $\sigma_2 D(\mathbf{k})$, Langer's
expression reduces to Eq.~(\ref{sigma.like.Langer}). It is also
evident from Eq.~(\ref{Lambda.Langer}) that for the
momentum-independent self-energy the bubble conductivity coincides
with the conductivity calculated in the {\em bare bubble\/}
approximation, but clearly this is {\em not the case\/} of the DDW
gap. Expression (\ref{sigma.like.Langer}) can also be derived in
the lowest order approximation from a general expression for the
dc conductivity, with the vertex corrections taken into account,
obtained by Eliashberg \cite{Eliashberg:1962:JETP}.

Making the nodal approximation
(see Eq.~(\ref{bubble.final})) we get
\begin{equation}
\label{conductivity.general.final}
\begin{split}
\sigma(B,T) & = \frac{2 \pi e^2}{v_{F}v_{D}}
\int \frac{d^{2}p}{(2\pi)^{2}} \int \limits_{-\infty}^{\infty} d\omega
(-n_F^{\prime}(\omega-\mu)) \\
& \times \left( v_F^2 \mbox{tr} \left[ A_D(\omega, {\bf p}) \gamma^{1}
A_D(\omega, {\bf p}) \gamma^{1}  \right] +
v_D^2 \mbox{tr} \left[ A_D(\omega, {\bf p}) \gamma^{2}
A_D(\omega, {\bf p}) \gamma^{2}  \right] \right).
\end{split}
\end{equation}

Finally we note that one could also calculate conductivity
directly working with the nodal Lagrangian (\ref{Dirac.conjugated.Lagrangian})
and expressing the bubble in terms of the corresponding electrical current operator
\begin{equation}
j_x = e v_{F} \bar{\chi}_s \gamma^{1} \chi_s, \qquad j_y = e v_{D}
\bar{\chi}_s \gamma^{2} \chi_s,
\end{equation}
which is also valid in the presence of an external field.
This way of calculation is  exactly the same as in Ref.~\cite{Yang:2002:PRB}
and final result agrees with (\ref{conductivity.general.final}).

\subsection{Calculation of conductivity}
\label{sec:conductivity.calculation}

Straightforward calculation of the trace in
Eq.~(\ref{conductivity.general.final}), with $A_D(\omega,
\mathbf{k})$ from Eqs.~(\ref{AD.clean}) and (\ref{f}) but with the
$\delta$ functions in Eq.~(\ref{AD.clean}) replaced by the
Lorentzians (\ref{Gamma}), gives
\begin{equation}
\label{tr.same}
\begin{split}
& \mbox{tr} \left[ A_D(\omega, {\bf p}) \gamma^{1,2}
A_D(\omega, {\bf p}) \gamma^{1,2}  \right] =
 \frac{4 \Gamma^2}{\pi^2} \exp \left( - \frac{2 \mathbf{p}^2}{eB} \right) \\
& \times
\sum_{n,m =0}^{\infty} (-1)^{n+m+1}
\frac{[(\omega^2 + M_n^2 + \Gamma^2) (\omega^2 + M_m^2 + \Gamma^2) - 4
\omega^2 \Delta^2][L_n L_{m-1} + L_{n-1} L_{m}] \mp
32 (p_1^2 - p_2^2) L_{n-1}^1 L_{m-1}^1}
{[(\omega^2 + M_n^2 + \Gamma^2)^2 - 4 \omega^2 M_n^2]
[(\omega^2 + M_m^2 + \Gamma^2)^2 - 4 \omega^2 M_m^2]},
\end{split}
\end{equation}
where all Laguerre's polynomials depend on $\frac{2
\mathbf{p}^2}{eB}$, and the minus sign corresponds to $\gamma^1$
and the plus sign to $\gamma^2$, respectively.

The integration over momentum $p$ can be easily done after extending the upper limit
of integration to $\infty$, so that one can use
the orthogonality of Laguerre's polynomials
\begin{equation}
\int \limits_{0}^{\infty} d xe^{-x} x^{\alpha} L_{m}^{\alpha} (x) L_{n}^{\alpha} (x)
= \Gamma(1+\alpha) \frac{(n+\alpha)!}{n! \alpha!} \delta_{mn}
\quad \mbox{with} \quad x = \frac{2 \mathbf{p}^2}{e B},
\end{equation}
and obtain electrical conductivity in terms of
the sum over the transitions between neighboring Landau levels
\begin{equation}
\label{sigma.final1}
\begin{split}
\sigma(B,T) & =  e^2 \frac{v_F^2 + v_D^2}{v_F v_D}
\frac{e B \Gamma^2}{2 \pi^2 T} \\
& \times \sum_{n=0}^{\infty}
\int \limits_{-\infty}^{\infty} d\omega
\frac{1}{\cosh^2 \frac{\omega - \mu}{2T}}
\frac{(\omega^2 + M_n^2 + \Gamma^2) (\omega^2 + M_{n+1}^2 + \Gamma^2) - 4
\omega^2 \Delta^2}
{[(\omega^2 + M_n^2 + \Gamma^2)^2 - 4 \omega^2 M_n^2]
[(\omega^2 + M_{n+1}^2 + \Gamma^2)^2 - 4 \omega^2 M_{n+1}^2]}.
\end{split}
\end{equation}
The sum over $n$ in Eq.~(\ref{sigma.final1}) can be expressed via
the digamma function $\psi$  as described in
Ref.~\cite{Ferrer:2002}, and the final expression for the
electrical conductivity is
\begin{equation}
\label{sigma.final2}
\sigma =  e^2  \alpha
\int \limits_{-\infty}^{\infty} \frac{d \omega}{4T\cosh^2 \frac{\omega - \mu}{2T}}
\mathcal{A}(\omega,B,\Gamma,\Delta(B)),
\end{equation}
where we introduced the function $\mathcal{A}$
\begin{equation}
\label{A.def}
\begin{split}
\mathcal{A}(\omega,B,\Gamma,\Delta(B))=
\frac{1}{\pi^2}
\frac{\Gamma^2}{(e B)^2 + (2 \omega \Gamma)^2}  & \left\{ 2 \omega^2 +
\frac{(\omega^2 + \Delta^2 + \Gamma^2)(eB)^2 - 2 \omega^2 (\omega^2 - \Delta^2 + \Gamma^2) eB}
{(\omega^2 - \Delta^2 - \Gamma^2)^2 + 4 \omega^2 \Gamma^2} \right. \\
&\left. -
\frac{\omega(\omega^2 - \Delta^2 + \Gamma^2)}{\Gamma} \mbox{Im}
\psi \left( \frac{\Delta^2 + \Gamma^2 - \omega^2 - 2 i \omega \Gamma}
{2eB}\right) \right\}
\end{split}
\end{equation}
and
\begin{equation}
\label{alpha}
\alpha = \frac{v_F}{v_D} + \frac{v_D}{v_F}.
\end{equation}

Another representation of Eq.~(\ref{A.def}) can be obtained using
the series representation of $\psi$ function and writing the
expression in curly brackets in fractions of $1/(\Gamma^2 + x^2)$,
\begin{equation}
\label{A.series}
\begin{split}
\mathcal{A}(\omega,B,\Gamma,\Delta(B))=
\frac{1}{\pi^2}
\frac{\Gamma^2}{(e B)^2 + (2 \omega \Gamma)^2}  & \left\{ 2 \omega^2 +
\frac{\frac{(eB)^2}{2} + eB \omega(\omega + \Delta)}
{(\omega+\Delta)^2 + \Gamma^2} +
\frac{\frac{(eB)^2}{2} + eB \omega(\omega - \Delta)}
{(\omega-\Delta)^2 + \Gamma^2}
\right. \\
&\left. + eB\omega \sum_{n=1}^{\infty} \frac{1}{M_n}
\left[\frac{\Delta^2 + M_n^2 + 2 \omega M_n}{(\omega + M_n)^2 + \Gamma^2} +
\frac{2 \omega M_n - \Delta^2 - M_n^2}{(\omega - M_n)^2 + \Gamma^2}.
\right] \right\},
\end{split}
\end{equation}
The representation (\ref{A.series}) is particularly convenient for
studying the narrow width limit $\Gamma \ll T, \sqrt{eB}$ when we
can replace the fractions $\Gamma/(\Gamma^2 + x^2)$ by $\pi
\delta(x)$:
\begin{equation}
\label{A.series.delta}
\begin{split}
\mathcal{A}(\omega,B,\Gamma,\Delta(B))=
\frac{\Gamma}{\pi} &  \left\{
\frac{1}{(e B)^2 + 4 \Delta^2 \Gamma^2}
\left[\frac{(eB)^2}{2} \delta(\omega + \Delta) +
\frac{(eB)^2}{2} \delta(\omega - \Delta) \right] \right. \\
& \left.  + \sum_{n=1}^{\infty}
\frac{2 (e B)^2 n}{(e B)^2 + 4 (\Delta^2 + 2 eB n) \Gamma^2}
\left[ \delta(\omega + M_n) + \delta(\omega - M_n) \right] \right\},
\end{split}
\end{equation}
where we kept $\Gamma^2$ in the denominators in order to be able
to reproduce a smooth behavior of $\sigma(B)$ and $\kappa(B)$ in
the limit $B \to 0$. We are now in a position to study different
asymptotic regimes defined by different relations among the
parameters $\Gamma$, $T$, $\mu$, $B$, and $\Delta$.

\subsection{Zero magnetic field}
\label{sec:conductivity.B=0}

We begin with the limit of vanishing magnetic field ($B =0$).
Using the large $z$ asymptote of the $\psi$ function
\begin{equation}
\psi(z) = \ln z - \frac{1}{2z} - \frac{1}{12z^2} + \frac{1}{120z^4} +
O\left(\frac{1}{z^6} \right)
\end{equation}
we obtain
\begin{equation}
\label{A.B=0}
\mathcal{A}(\omega,B=0,\Gamma,\Delta)=
\frac{1}{2 \pi^2}
\left[ 1 + \frac{\omega^2 - \Delta^2 + \Gamma^2}{2 |\omega| \Gamma}
\left(\frac{\pi}{2} - \arctan \frac{\Delta^2 + \Gamma^2 - \omega^2}
{2|\omega| \Gamma}
\right) \right].
\end{equation}
Putting $\Delta = 0$ also in Eq.~(\ref{A.B=0}), we get
\begin{equation}
\label{A.B=0.Delta=0}
\mathcal{A}(\omega,B=0,\Gamma,\Delta =0)=
\frac{1}{2 \pi^2}
\left[ 1 + \frac{\omega^2  + \Gamma^2}{\omega \Gamma}
\arctan \frac{\omega}{ \Gamma} \right].
\end{equation}

\subsubsection{Limit $T \to 0$}

The limit $T \to 0$ is significantly simplified by the fact that
$\cosh^{-2}$ term in Eq.~(\ref{sigma.final2}) can be replaced by
the $\delta$ function
\begin{equation}
\label{electric.B=0.final}
\sigma  =  e^2  \alpha \mathcal{A}(\mu,B=0,\Gamma,\Delta) ,
\end{equation}
where $\mathcal{A}$ is given by Eq.~(\ref{A.B=0}). For the case of
zero gap $\Delta =0$ using Eq.~(\ref{A.B=0.Delta=0}) we obtain
that for $\mu =0$
\begin{equation}
\label{sigma.mu=0}
\sigma (\mu =0)=  \frac{e^2 \alpha}{\pi^2}
\end{equation}
and for $|\mu| \gg \Gamma$
\begin{equation}
\label{sigma.mu>Gamma}
\sigma =  \frac{e^2  \alpha}{4\pi} \frac{|\mu|}{\Gamma}.
\end{equation}
Comparing Eq.~(\ref{sigma.mu=0}) with Eq.~(57) of YN, one can see
that our result is twice as large, but contains the same prefactor
$\alpha$ that reflects the fact that the electrical current
operator has the component $\sim \mathbf{v}_D$. On the other hand,
the result of KC is $\sim v_F/v_D$ due to the fact that they took
the current without the $\mathbf{v}_D$ component, and it is twice
as large as Eq.~(\ref{sigma.mu=0}) because KC integrated over the
full Brillouin zone instead of the reduced one. Expression
(\ref{sigma.mu=0}) can also be compared with its superconducting
counterpart \cite{Durst:2000:PRB} $\sigma_{SC} = (e^2/\pi^2)
v_{F}/v_{\Delta}$, where $v_{\Delta}$ is the $d$SC gap velocity.
As one can see the numerical prefactor is {\em exactly the
same\/}, but there is no $v_{\Delta}/v_{F}$ term in $\sigma_{SC}$
because the electrical current in the $d$SC state is
$\sim\mathbf{v}_F$.

Finally one can also compare our $\mu \neq 0$ expression (\ref{sigma.mu>Gamma})
with Eq.~(13) of KC:
\begin{equation}
\label{KC.mu}
\sigma(T \ll D_0) \simeq \frac{e^2}{2 \pi} \frac{v_F}{v_D} \frac{D_0}{\gamma_0},
\end{equation}
where in the Born limit $\gamma_0$ is related to $\Gamma(\omega)$
via $\Gamma (\omega) = \gamma_0 [(\omega + \mu)/D_0]$.
Substituting the value of $\Gamma(0)$ in
Eq.~(\ref{sigma.mu>Gamma}) one can see that it reduces to
Eq.~(\ref{KC.mu}) except for the abovementioned differences in the
factors.

\subsubsection{Limit $T \ll \Gamma$}

In order to obtain low-temperature corrections to  conductivities
it is convenient to apply the Sommerfeld expansion
\begin{equation}
\label{Fermi.expansion}
\begin{split}
& \int \limits_{-\infty}^{\infty}
\frac{(\omega - \mu)^n}{4T\cosh^2 \frac{\omega - \mu}{2T}} f(\omega) d \omega \simeq
\int \limits_{-\infty}^{\infty}
\frac{(\omega - \mu)^n}{4T\cosh^2 \frac{\omega - \mu}{2T}} d \omega f(\mu) \\
& +
\int \limits_{-\infty}^{\infty}
\frac{(\omega - \mu)^{n+1}}{4T\cosh^2 \frac{\omega - \mu}{2T}} d \omega f^{\prime}(\mu) +
\frac{1}{2}
\int \limits_{-\infty}^{\infty}
\frac{(\omega - \mu)^{n+2}}{4T\cosh^2 \frac{\omega - \mu}{2T}} d \omega
f^{\prime \prime}(\mu) +
O \left( \left(\frac{T}{\mbox{max} (\mu, \Gamma)} \right)^{n+3} \right).
\end{split}
\end{equation}
Here the function $f$ should be nonsingular and not too rapidly
varying in the vicinity of $\omega = \mu$ and the integrals on the
right hand side of Eq.~(\ref{Fermi.expansion}) are
\begin{equation}
\int \limits_{-\infty}^{\infty} \frac{d \omega}{4T\cosh^2 \frac{\omega - \mu}{2T}} =1, \qquad
\int \limits_{-\infty}^{\infty}
\frac{d \omega (\omega - \mu)^2}{4T\cosh^2 \frac{\omega - \mu}{2T}} = \frac{\pi^2 T^2}{3},
\qquad
\int \limits_{-\infty}^{\infty}
\frac{d \omega(\omega - \mu)^4}{4T\cosh^2 \frac{\omega - \mu}{2T}} = \frac{7 \pi^4 T^4}{15} .
\end{equation}
It is easy to derive the next to the leading term in $T^2$ to the
conductivity (\ref{electric.B=0.final}) for $\Delta =0$,
\begin{equation}
\label{sigma.B=0.final2}
\sigma = e^2 \alpha \left[ \mathcal{A}(\omega = \mu, 0, \Gamma, 0) +
\frac{\pi^2 T^2}{6}  \mathcal{A}^{\prime \prime}_\omega(\omega= \mu, 0, \Gamma,0)
\right] ,
\end{equation}
where the derivatives of $\mathcal{A}$ with respect to $\omega$ are
\begin{equation}
\label{A.derivatives}
\begin{split}
& \mathcal{A}^{\prime}_{\omega} (\omega, 0, \Gamma, 0) = \frac{1}{2 \pi^2 \omega}
\left( 1 + \frac{\omega^2 - \Gamma^2}{\omega \Gamma} \arctan \frac{\omega}{\Gamma} \right), \\
& \mathcal{A}^{\prime \prime}_{\omega} (\omega, 0, \Gamma, 0) =  \frac{1}{\pi^2 \omega^2}
\left(\frac{\Gamma}{\omega} \arctan \frac{\omega}{\Gamma} - \frac{\Gamma^2}{\omega^2 + \Gamma^2}
\right).
\end{split}
\end{equation}
For $\mu = 0$ case Eq.~(\ref{sigma.B=0.final2}) gives the $T^2$ correction to
the expression (\ref{sigma.mu=0}):
\begin{equation}
\sigma (\mu =0)=  \frac{e^2  \alpha}{\pi^2}
\left[1 + \frac{\pi^2}{9} \frac{T^2}{\Gamma^2} \right], \qquad T \ll \Gamma.
\end{equation}

\subsubsection{Limit $T \gg \Gamma$}
For $\Gamma \to 0$ from Eq.~(\ref{A.B=0}) we get
\begin{equation}
\label{A.B=0.T>Gamma}
\mathcal{A}(\omega,B=0,\Gamma,\Delta)=
\frac{1}{4 \pi^2}
\frac{\omega^2 - \Delta^2}{|\omega| \Gamma}
\pi \theta( \omega^2 - \Delta^2).
\end{equation}
For the $\Delta = 0$ case retaining more terms in the expansion of
Eq.~(\ref{A.B=0.Delta=0}), we obtain
\begin{equation}
\label{A.B=D=0.T>Gamma}
\mathcal{A}(\omega,B=0,\Gamma,\Delta =0)=
\frac{1}{4 \pi}  \left( \frac{|\omega|}{\Gamma} +
\frac{\Gamma}{|\omega|} \right).
\end{equation}
Substituting Eq.~(\ref{A.B=D=0.T>Gamma}) in the expression
(\ref{sigma.final2}) for $\sigma$ we arrive at
\begin{equation}
\label{sigma.T>Gamma}
\sigma  =  e^2  \alpha
\frac{T}{2\pi \Gamma}  \ln \left(2 \cosh
\frac{\mu}{2T} \right),  \quad T \gg \Gamma .
\end{equation}
As one can easily see for $\mu \gg T$ Eq.~(\ref{sigma.T>Gamma}) reduces to
Eq.~(\ref{sigma.mu>Gamma}) reflecting the fact that when $\mu$ is the largest parameter
the value of conductivity is not sensitive to the relation between $T$ and $\Gamma$.

As mentioned before, the magnetic catalysis phenomenon
\cite{Gusynin:1995:PRD,Ferrer:2002,Gorbar:2002:PRB} implies that
the gap $\Delta$ is generated {\em only\/} in the presence of an
external field. Nevertheless, one can also study the consequences
of the gap opening even for $B =0$ case to gain a deeper insight
to a more complicated case $B, \Delta \neq 0$. In particular,
considering the $\Delta \gg T$ case, we obtain from
Eqs.~(\ref{A.B=0.T>Gamma}) and (\ref{sigma.final2}) that
\begin{equation}
\label{sigma.Delta}
\sigma  =  \frac{e^2  \alpha }{16\pi
T \Gamma} \int \limits_{\Delta}^{\infty} d \omega
\frac{\omega^2 - \Delta^2}{\omega} \left[\frac{1}{\cosh^2 \frac{\omega-\mu}{2T}}
+ (\mu \to - \mu) \right] \simeq
\frac{e^2 \alpha}{4 \pi \Gamma}
\begin{cases}
& \dfrac{\mu^2 - \Delta^2}{|\mu|} - \dfrac{\pi^2 T^2 \Delta^2}{3 \mu^2 |\mu|}, \qquad \quad
|\mu| > \Delta, \\
& 2 T \ln 2, \qquad \qquad \qquad \qquad \quad |\mu| = \Delta,\\
& 4 T \cosh \dfrac{\mu}{T} \exp\left( - \dfrac{\Delta}{T} \right), \qquad
|\mu| < \Delta.
\end{cases}
\end{equation}
It is clear from Eq.~(\ref{sigma.Delta}) that the opening of the gap
$\Delta$ results in the thermally activated behavior of conductivity
only for $|\mu | < \Delta$. As discussed in
Sec.~\ref{sec:sigma.numerical} this observation remains valid even in the
presence of an external field.

\subsection{Nonzero magnetic field}

There are not so many cases available for analytical investigation
for $B \neq 0$ and we have to integrate numerically
Eq.~(\ref{sigma.final2}) with $\mathcal{A}$ given by
Eq.~(\ref{A.def}). Nevertheless in a few cases analytical
expressions for the conductivity can be obtained and here we begin
with considering these results.

\subsubsection{Limit $T \to 0$}

As one can notice, Eq.~(\ref{electric.B=0.final}) is in fact valid
even for nonzero $B$ because only the first of term of the
Sommerfeld expansion (\ref{Fermi.expansion}) contributes, i.e.,
\begin{equation}
\label{electric.B.final}
\sigma  =  e^2  \alpha \mathcal{A}(\mu,B,\Gamma,\Delta) ,
\qquad \forall \mu, \quad \Gamma \neq 0.
\end{equation}

\subsubsection{Narrow width case}

To study the narrow width limit $\Gamma \to 0$ we use the
representation (\ref{A.series.delta}) to arrive at
\begin{equation}
\begin{split}
\label{sigma.field.narrow}
\sigma  =  e^2  \alpha \frac{\Gamma}{4 \pi T}
& \left\{
\frac{(eB)^2}{2[(e B)^2 + 4 \Delta^2 \Gamma^2]}
\left[\frac{1}{\cosh^2 \frac{\Delta + \mu}{2T}} +
\frac{1}{\cosh^2 \frac{\Delta - \mu}{2T}} \right] \right. \\
& \left.  + \sum_{n=1}^{\infty}
\frac{2 (e B)^2 n}{(e B)^2 + 4 (\Delta^2 + 2 eB n) \Gamma^2}
\left[ \frac{1}{\cosh^2 \frac{\sqrt{\Delta^2 + 2 eBn}+\mu}{2T}} +
  \frac{1}{\cosh^2 \frac{\sqrt{\Delta^2+ 2 eBn}-\mu}{2T}}
\right] \right\}.
\end{split}
\end{equation}
The fact that the conductivity is proportional to the scattering
rate $\Gamma$ in the limit $\Gamma \to 0$ means that in contrast
to the zero-field case (see Sec.~\ref{sec:conductivity.B=0}) it
results from transitions of quasiparticles between neighboring
cyclotron orbits.

Using Eq.~(\ref{sigma.field.narrow}) one easily gets the
asymptotic expression for conductivity in the {\em strong-field
limit}, $\sqrt{e B} \gtrsim 4 T$ for $\mu = \Delta =0$
\begin{equation}
\label{sigma.Bstrong}
\sigma  =  e^2  \alpha \frac{\Gamma}{4 \pi T}
\end{equation}
that shows that $\sigma$ becomes field independent in the strong field.

\subsubsection{Numerical calculation of electrical conductivity}
\label{sec:sigma.numerical}

To investigate numerically the behavior of electrical and thermal
conductivities, and to make comparison with experiment, we need to
restore all model parameters, such as $\hbar$, $c$, $k_B$,
$v_{F}$, and $v_D$ in Eqs.~(\ref{sigma.final2}) and (\ref{A.def}).
As discussed after Eq.~(\ref{scaling}) the prefactor $\alpha$ in
Eq.~(\ref{sigma.final2}) is already fixed and one should only
substitute $T \to k_B T$, $\Gamma \to \hbar \Gamma$, and $e B \to
(\hbar v_F v_D/c) e B$. It is convenient to measure all energetic
quantities in  K, which results in the following replacement:
\begin{equation}
\label{units}
\sqrt{2 e B} \to \sqrt{\frac{\hbar v_F v_D 2 e B}{c}} [\mbox{K}] = 63.9
\sqrt{\frac{v_{D}}{v_{F}}} v_{F} [\mbox{eV} \cdot \mbox{\AA}] \sqrt{B [\mbox{Tesla}]} =
4.206 \times 10^{-6}
\sqrt{\frac{v_{D}}{v_{F}}} v_{F} [\mbox{cm}/\mbox{s}] \sqrt{B [\mbox{Tesla}]} ,
\end{equation}
where in the first equality $v_{F}$ is given in $\mbox{eV} \cdot
\mbox{\AA}$ and in $\mbox{cm}/\mbox{s}$ in the second one. In
particular, for $v_F = 1.5 \mbox{eV} \cdot \mbox{\AA}$ [this
roughly agrees with the value of $t$ given after
Eq.~(\ref{Green.nodal})] and $v_F/v_D = 24$ using
Eq.~(\ref{units}) we obtain that $e B \to 200 \cdot \mbox{K}^2
\cdot B [\mbox{Tesla}]$. In what follows we use this estimate to
compute all numerical expressions. There is a larger uncertainty
for the value of $\Gamma$ that we could choose for our
computations. For example, for clean YBCO monocrystals the
estimated \cite{Ando:2000:PRB} value of the scattering rate due to
impurities $\Gamma_0 \sim 1-2 \mbox{K}$, so that we use the value
$\Gamma = 2 \mbox{K}$.

\begin{figure}[h]
\centering{
\includegraphics[width=8cm]{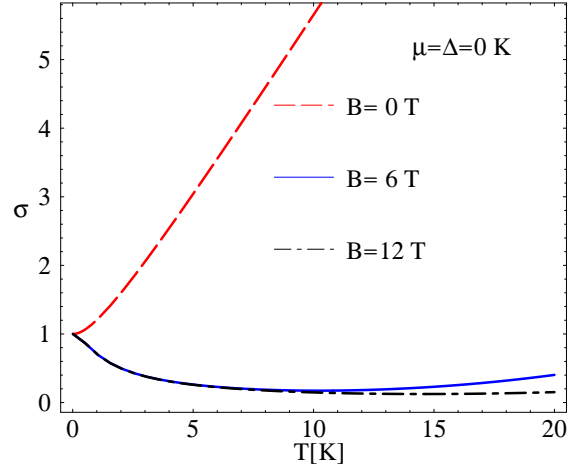}}
\caption{The normalized conductivity $\sigma/\sigma(\mu =T = B =0)$
as function of temperature, $T$, for three different values of magnetic
field $B$ at half-filling, $\mu= 0$. The value of $\sigma(\mu =T = B =0)$
is given by Eq.~(\ref{sigma.mu=0}).}
\label{fig:3}
\end{figure}

\begin{figure}[h]
\centering{
\includegraphics[width=8cm]{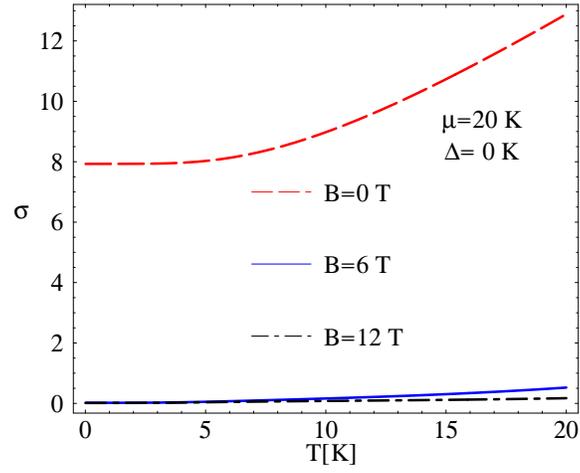}}
\caption{The normalized conductivity $\sigma/\sigma(\mu =T = B =0)$
as function of temperature, $T$, for three different values of magnetic
field $B$ away from half-filling, $\mu= 20 \mbox{K}$.}
\label{fig:4}
\end{figure}

In Figs.~\ref{fig:3} and \ref{fig:4} we show the temperature
dependence of the conductivity for three different values of the
applied field at half-filling (Fig.~\ref{fig:3}) and slightly away
from it (Fig.~\ref{fig:4}). For $\mu = B= 0$ the conductivity
increases as the temperature $T$ grows. However, when the magnetic
field is nonzero, the dependence $\sigma(T)$ becomes non-monotonic
and there is first a decrease in $\sigma(T)$ as the temperature
increases. For $\mu, B \neq 0$ (see Fig.~\ref{fig:4}) the
conductivity slowly increases as the temperature grows and this
dependence becomes almost flat as the field increases. This
tendency  agrees with Eqs.~(\ref{sigma.field.narrow}) and
(\ref{sigma.Bstrong}). In zero field both figures reveal an
```insulator'' (i.e., increasing with temperature) behavior in
agreement with our analytical result (\ref{sigma.T>Gamma}). This
behavior is a consequence of using a constant value for the
scattering rate $\Gamma$ in our model. The growth of the
conductivity with increasing temperature is directly related to
the increasing number of thermally excited quasiparticles. The
choice of temperature-independent $\Gamma$ might be reasonable in
the narrow low-temperature region. It is essential, however, that
the observed decrease and flattening $\sigma(T,B)$ in the nonzero
external field are due to the assumption $\Gamma(B) =
\mbox{const}$ discussed after Eq.~(\ref{Gamma}) and are not
related to the $\Gamma(T) = \mbox{const}$ approximation just
mentioned.

Comparing Figs.~\ref{fig:3} and \ref{fig:4},  one can see that in
zero field the increase of $|\mu|$ (opening of the pockets on the
Fermi surface) leads to the increase of the conductivity. On the
contrary, in the presence of the external field, the conductivity
{\em decreases} as the pockets on the Fermi surface open. One can
notice that the value of conductivity $\sigma$ at $\mu = T = 0$
{\em is field independent\/}. This is due to the fact that for
$\omega = 0$, the function $\mathcal{A}$ [see Eq.~(\ref{A.def})
and Fig.~\ref{fig:14}] becomes field independent.

Using Eqs.~(\ref{sigma.final2}) and (\ref{A.def}) it is also
possible to investigate the case of $\Delta \neq 0$. The main
results for $B \neq 0$ can be foreseen from
Eq.~(\ref{sigma.Delta}) and are the following. When $T,|\mu|
\lesssim \Delta$ the behavior of $\sigma$ becomes thermally
activated, i.e., governed by the factor $\sim \exp(-\Delta/T)$. In
contrast, for $\Delta \lesssim |\mu|$ the behavior of $\sigma(T)$
is rather similar to the case of $\Delta = 0$. This reflects the
fact that the gap $\Delta$ induced by the magnetic catalysis, in
contrast to a superconducting gap, {\em is not tied} to the Fermi
surface and there are gapless excitations for $|\mu| \gtrsim
\Delta$.

\begin{figure}[h]
\centering{
\includegraphics[width=8cm]{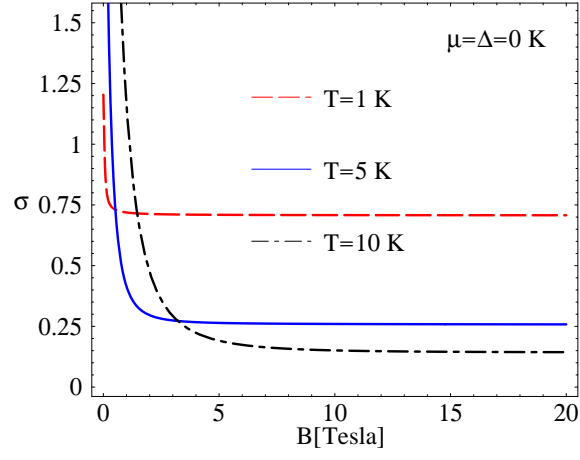}}
\caption{The normalized conductivity $\sigma/\sigma(\mu =T = B
=0)$ as function of field, $B$, for three different values of
temperature $T$ at half-filling, $\mu= 0$.}
\label{fig:5}
\end{figure}

\begin{figure}[h]
\centering{
\includegraphics[width=8cm]{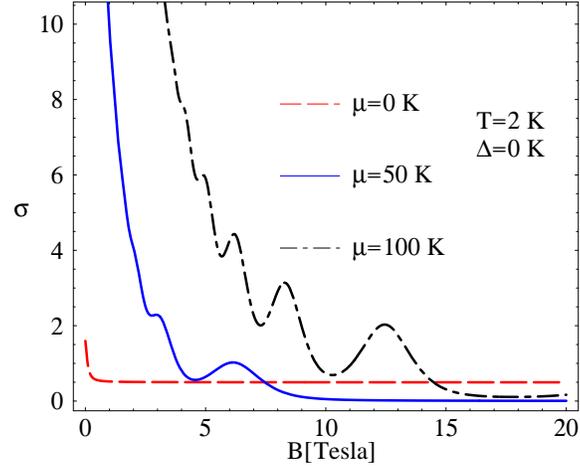}}
\caption{The normalized conductivity $\sigma/\sigma(\mu =T = B
=0)$ as function of field, $B$, for three different values of
chemical potential $\mu$ at $T = 2 \mbox{K}$.}
\label{fig:6}
\end{figure}

In Figs.~\ref{fig:5} and \ref{fig:6} we show, respectively, the
dependence of $\sigma(B)$ for three different values of $T$ and
fixed $\mu$ and for three different values of $\mu$ at fixed $T$.
Both figures show that $\sigma$ decreases as a function of $B$.
This decrease reflects the fact mentioned before
Eq.~(\ref{sigma.final1}) that in the presence of magnetic field
only transitions between neighboring Landau levels contribute into
electrical (and thermal) conductivity. A further increase of the
field and entering in the {\em strong field regime},  $\sqrt{e B}
\gtrsim 4 T$ suppresses the transition between Landau levels, so
that as one can see from  Fig.~\ref{fig:5}, the conductivity
$\sigma(B)$ becomes field independent. Furthermore for $T =10
\mbox{K}$ the value of normalized conductivity is $\approx 0.15$,
in agreement with Eq.~(\ref{sigma.Bstrong}), which is derived
under the assumption $\Gamma \ll T$ and taking into account only
the transition between the lowest levels. In Fig.~\ref{fig:6} we
observe Shubnikov-de Haas oscillations of the conductivity that
are due to the Landau-level crossing of the Fermi pockets (an
analytical treatment of the magnetic oscillations in the
conductivities will be presented elsewhere).

\section{Thermal conductivity}
\label{sec:thermal}

\subsection{General expression for thermal conductivity}

Thermal conductivity can be calculated from the
the energy current-current correlation function
\begin{equation}
\label{polarization.EE}
\Pi^{EE}_{\alpha \beta}(i \Omega) = - \int \limits_{0}^{\beta} d \tau e^{i \Omega \tau}
\langle T_{\tau} j_{\alpha}^{E \dagger}(\tau, \mathbf{0})
j_{\beta}^{E}(0, \mathbf{0}) \rangle,
\end{equation}
and the correlation function of energy current with the electrical current
\begin{equation}
\label{polarization.EC}
\Pi_{\alpha \beta}^{EC}(i \Omega) = -
\int \limits_{0}^{\beta} d \tau e^{i \Omega \tau}
\langle T_{\tau} j_{\alpha}^{E \dagger}(\tau, \mathbf{0})j_{\beta}(0, \mathbf{0}) \rangle
\end{equation}
using a thermal Kubo formula \cite{Mahan:book,foot1}
\begin{equation}
\label{thermal.def}
\frac{\kappa(\Omega)}{T} = - \frac{1}{T^2}
\frac{\mbox{Im} \Pi_{R}^{EE}(\Omega +i0)}{\Omega} - S^2(\Omega) \sigma(\Omega,B,T) =
- \frac{1}{T^2} \frac{\mbox{Im} \Pi_{R}^{EE}(\Omega +i0)}{\Omega} -
\frac{1}{T^2} \frac{[\mbox{Im} \Pi_{R}^{EC}(\Omega +i0)]^2}
{\sigma(\Omega) \Omega^2}.
\end{equation}
Here $\Pi_{R}^{EE}(\Omega +i0) = \Pi^{EE}(i \Omega \to \Omega +
i0)$ is the longitudinal polarization [see
Eq.~(\ref{tensor2scalar})] and $S(\Omega,B,T)$ is the thermopower
\begin{equation}
S(\Omega) = - \frac{1}{T} \frac{\mbox{Im} \Pi_{R}^{EC} (\Omega)}
{\mbox{Im} \Pi_{R}^{CC} (\Omega)},
\end{equation}
with the longitudinal $\Pi_{R}^{EC}(\Omega +i0) = \Pi^{EC}(i
\Omega \to \Omega + i0)$. The term with the thermal power $S$
ensures that the energy current is evaluated under the condition
of vanishing electrical current \cite{Langer:1962:PR} (see also
Ref.~\cite{Mahan:book}). Usually, for $T \ll \mu $ this term is
considered to be unimportant because it is $ \sim T^2/\mu^2$ times
less than the first term of Eq.~(\ref{thermal.def}). It is also
zero for $\mu =0$, but nevertheless its contribution is important
for the case of interest, $|\mu| \lesssim T$.

The structure of the energy current (\ref{heat.current2.B=0}) is
similar to that of the electrical current
(\ref{electric.current.DDW}). Hence, the calculation of the
corresponding polarization functions (\ref{polarization.EE})  and
(\ref{polarization.EC}) is almost identical to calculating  the
bubble (\ref{el.cur-cur.tensor}). Using the general result
(\ref{bubble.final}) to the case of interest: $g = g^{\prime} =
\omega+ \Omega/2 - \mu$, we arrive at
\begin{equation}
\label{thermal.conductivity.general.final1}
\begin{split}
- \frac{\mbox{Im} \Pi_R^{EE}(\Omega)} {T^2 \Omega}& = \frac{\pi}{v_{F}v_{D}}
\int \frac{d^{2}p}{(2\pi)^{2}} \int \limits_{-\infty}^{\infty} d\omega
\frac{\tanh \frac{\omega - \mu + \Omega}{2T} - \tanh \frac{\omega - \mu}{2T}}{\Omega}
\left( \frac{\omega - \mu + \Omega/2 }{T} \right)^2 \\
& \times \left[ v_F^2 \mbox{tr} \left[ A_D(\omega, {\bf p}) \gamma^{1}
A_D(\omega + \Omega, {\bf p}) \gamma^{1}  \right] +
v_D^2 \mbox{tr} \left[ A_D(\omega, {\bf p}) \gamma^{2}
A_D(\omega + \Omega, {\bf p}) \gamma^{2}  \right] \right],
\end{split}
\end{equation}
and for
$g = \omega+ \Omega/2 - \mu$, $g^{\prime} = e$
\begin{equation}
\label{thermal.conductivity.general.final2}
\begin{split}
- \frac{\mbox{Im} \Pi_R^{EC}(\Omega)} {T \Omega}& = \frac{e \pi}{v_{F}v_{D}}
\int \frac{d^{2}p}{(2\pi)^{2}} \int \limits_{-\infty}^{\infty} d\omega
\frac{\tanh \frac{\omega - \mu + \Omega}{2T} - \tanh \frac{\omega - \mu}{2T}}{\Omega}
\frac{\omega - \mu + \Omega/2 }{T} \\
& \times \left[ v_F^2 \mbox{tr} \left[ A_D(\omega, {\bf p}) \gamma^{1}
A_D(\omega + \Omega, {\bf p}) \gamma^{1}  \right] +
v_D^2 \mbox{tr} \left[ A_D(\omega, {\bf p}) \gamma^{2}
A_D(\omega + \Omega, {\bf p}) \gamma^{2}  \right] \right].
\end{split}
\end{equation}
Similarly to the case of the electrical current, one could also
obtain the energy current operator from the nodal Lagrangian
(\ref{Dirac.conjugated.Lagrangian}) in the external magnetic field
as done in Ref.~\cite{Ferrer:2002},
\begin{equation}
\label{energy.current.nodal}
j_x^E =  \frac{i v_{F}}{2} ( \bar{\chi}_s \gamma^{1} \partial_t \chi_s -
\partial_t \bar{\chi}_s \gamma^{1}  \chi_s),
\qquad
j_y^E =  \frac{i v_{D}}{2} ( \bar{\chi}_s \gamma^{2} \partial_t \chi_s -
\partial_t \bar{\chi}_s \gamma^{2}  \chi_s)
\end{equation}
to derive the expression (\ref{thermal.conductivity.general.final2}).
The current (\ref{energy.current.nodal})
corresponds to the current used by YN.

The calculation of thermal conductivity from
Eq.~(\ref{thermal.conductivity.general.final2})
follows exactly the same route as for
the electrical conductivity in Sec.~\ref{sec:conductivity.calculation} and
finally we arrive at
\begin{equation}
\label{thermal.conductivity.general.final3}
\begin{split}
\frac{\kappa (B,T)}{T} = \alpha & \left\{
\int \limits_{-\infty}^{\infty} d \omega \left( \frac{\omega-\mu}{T} \right)^2
\frac{1}{4T\cosh^2 \frac{\omega - \mu}{2T}}
\mathcal{A}(\omega,B,\Gamma,\Delta(B)) \right. \\
& - \left. \frac{e^2 \alpha}{\sigma (B,T)}
\left[ \int \limits_{-\infty}^{\infty} d \omega  \frac{\omega-\mu}{T}
\frac{1}{4T\cosh^2 \frac{\omega - \mu}{2T}}
\mathcal{A}(\omega,B,\Gamma,\Delta(B)) \right]^2 \right\} ,
\end{split}
\end{equation}
where $\mathcal{A}$ is the same function (\ref{A.def}) as for the
electrical conductivity and $\alpha$ is given by
Eq.~(\ref{alpha}). Thus we are ready to study the thermal
conductivity as a function of $\Gamma$, $T$, $\mu$, $B$, and
$\Delta$.

\subsection{Zero magnetic field}

To study the $B=0$ case we substitute  the expression for
$\mathcal{A}(B=0)$ given by Eq.~(\ref{A.B=0}) into
the general expression for thermal conductivity
(\ref{thermal.conductivity.general.final2}).

\subsubsection{Limit $T \to 0$}

Corresponding to Eq.~(\ref{electric.B=0.final}) limit $T\to 0$ is
given by
\begin{equation}
\label{kappa.B=0.final}
\frac{\kappa}{T} =  \frac{\pi^2}{3}
\alpha \mathcal{A}(\mu,B=0,\Gamma,\Delta) .
\end{equation}
Then using Eq.~(\ref{A.B=0.Delta=0}) for $\mu = \Delta =0$, we
obtain
\begin{equation}
\label{kappa.mu=0}
\frac{\kappa (\mu =0)}{T} = \frac{\alpha}{3} .
\end{equation}
The factor $\alpha$ in Eq.~(\ref{kappa.mu=0}) is present also in
the DDW \cite{Yang:2002:PRB,Kim:2002:PRB} and $dSC$
\cite{Durst:2000:PRB} cases where the same expression for the
thermal current has been used. The overall numerical factor $1/3$
in Eq.~(\ref{kappa.mu=0}) is the same as in
Ref.~\cite{Durst:2000:PRB} and it is different from
 Refs.~\cite{Yang:2002:PRB,Kim:2002:PRB} in the same way as the corresponding factor
in expression (\ref{sigma.mu=0})  for the electrical conductivity.

\subsubsection{Limit $T \ll \Gamma$}

Using the Sommerfeld expansion (\ref{Fermi.expansion}) we can also derive next to
the leading term in $T^2$ to the thermal conductivity for  $\Delta = 0$
\begin{equation}
\label{kappa.B=0.final2}
\frac{\kappa}{T} = \frac{\alpha}{3}\left[
\mathcal{A}(\omega = \mu,0,\Gamma, 0) + \frac{7 \pi^2 T^2}{30}
\mathcal{A}^{\prime \prime}_{\omega}(\omega = \mu,0,\Gamma, 0)
- \frac{\pi^2 T^2}{9}
\frac{(\mathcal{A}^{\prime}_{\omega}(\omega = \mu,0,\Gamma, 0))^2}
{\mathcal{A}(\omega = \mu,0,\Gamma, 0)}
\right],
\end{equation}
where the derivatives of $\mathcal{A}$ are given by Eq.~(\ref{A.derivatives}).
For $\mu = 0$, since there is no contribution from the thermal power term $\Pi^{EC}$,
Eq.~(\ref{kappa.B=0.final2}) reduces to the expression
\begin{equation}
\frac{\kappa (\mu =0)}{T} =
\frac{\alpha}{3} \left[ 1 + \frac{7 \pi^2}{15}
\frac{T^2}{\Gamma^2}\right] , \qquad T \ll \Gamma.
\end{equation}

\subsubsection{Limit $T \gg \Gamma$}

For $\Gamma \to 0$ and $\Delta = 0$ substituting
Eq.~(\ref{A.B=D=0.T>Gamma}) into the expression
(\ref{thermal.conductivity.general.final3}) for $\kappa$ we obtain
(see Appendix~\ref{sec:C})
\begin{equation}
\label{kappa.T>Gamma}
\begin{split}
\frac{\kappa}{T}  \simeq  \alpha \left[ \frac{\pi}{12}
\frac{\mu}{\Gamma} - \frac{\pi^3}{36} \frac{T^2}{\Gamma \mu}
\right], \qquad \mu \gg T
\end{split}
\end{equation}
where the first term $\sim \mu$  of $\kappa$ arises from the first
term of Eq.~(\ref{thermal.conductivity.general.final3}) and the
second term $\sim 1/\mu$ originates from the thermal power term
$\Pi^{EC}$. Thus, as one would expect, the thermal power
contribution into the thermal conductivity is  important only for
$|\mu| \lesssim T$.

For the case $\mu = \Delta = 0$ using Eq.~(\ref{A.B=D=0.T>Gamma})
we arrive at the expression
\begin{equation}
\label{kappa.T>Gamma.mu=0}
\begin{split}
\frac{\kappa(\mu =0)}{T}  = \frac{\alpha}{\pi} \left[ \frac{9
\zeta(3)}{4} \frac{T}{\Gamma} + \frac{\ln 2}{2} \frac{\Gamma}{T}
\right], \qquad \Gamma \ll T
\end{split}
\end{equation}
where $\zeta (z)$ is the Riemann zeta function. Finally using the
representation (\ref{A.B=0.T>Gamma}), which is valid for $\mu =0$
we can get the thermal conductivity for nonzero $\Delta$
\cite{Ferrer:2002}:
\begin{equation}
\label{kappa.Delta}
\frac{\kappa(\mu =0)}{T} = \frac{\alpha}{8\pi T^3
\Gamma} \int \limits_{\Delta}^{\infty} d \omega \frac{|\omega| (\omega^2 -
\Delta^2)}{\cosh^2 \frac{\omega}{2T}} \simeq \alpha \frac{\Delta^2}{\pi
\Gamma T} \exp \left( - \frac{\Delta}{T} \right), \qquad  \Delta \gg
T.
\end{equation}

\subsection{Nonzero magnetic field}

As for the electrical conductivity there are not so many cases
available for analytical investigation for $B \neq 0$ and we have
to integrate numerically
Eq.~(\ref{thermal.conductivity.general.final3}) with $\mathcal{A}$
given by Eq.~(\ref{A.def}). There are still a few cases when the
analytical treatment is  possible and we begin by looking at them.

\subsubsection{Limit $T \to 0$}

As one can check,
Eq.~(\ref{kappa.B=0.final})  is  valid even for
nonzero $B$, so that
\begin{equation}
\label{kappa.B.final}
\frac{\kappa}{T} =  \frac{\pi^2}{3}
\alpha \mathcal{A}(\mu,B=0,\Gamma,\Delta) , \qquad \forall \mu, \quad
\Gamma \neq 0.
\end{equation}

\subsubsection{Narrow width case}

Using the narrow width representation (\ref{A.series.delta}) for
$\mathcal{A}$, we arrive at the following result,
\begin{equation}
\label{kappa.field.narrow}
\begin{split}
\frac{\kappa}{T}  =   \alpha \frac{\Gamma}{4 \pi T^3}
& \left\{
\frac{(eB)^2}{2[(e B)^2 + 4 \Delta^2 \Gamma^2]}
\left[\frac{(\Delta + \mu)^2}{\cosh^2 \frac{\Delta + \mu}{2T}} +
\frac{(\Delta - \mu)^2}
{\cosh^2 \frac{\Delta - \mu}{2T}} \right] \right. \\
& \left.  + \sum_{n=1}^{\infty}
\frac{2 (e B)^2 n}{(e B)^2 + 4 (\Delta^2 + 2 eB n) \Gamma^2}
\left[ \frac{(\sqrt{\Delta^2 + 2 eBn}+\mu)^2}
{\cosh^2 \frac{\sqrt{\Delta^2 + 2 eBn}+\mu}{2T}} +
\frac{(\sqrt{\Delta^2+ 2 eBn}-\mu)^2}
{\cosh^2 \frac{\sqrt{\Delta^2+ 2 eBn}-\mu}{2T}}
\right] \right\} + \frac{\kappa_1}{T},
\end{split}
\end{equation}
where $\kappa_1$ is a term originating from the condition of
vanishing electrical current. As was discussed above, it is zero
for $\mu =0$ and can be neglected for $|\mu| \gg T$. In
particular, for $\mu = \Delta =0$ in the {\em strong field limit},
$\sqrt{e B} \gtrsim 4T$ we obtain from
Eq.~(\ref{kappa.field.narrow}) the expression
\begin{equation}
\label{kappa.Bstrong}
\frac{\kappa}{T}  =  \alpha \frac{8 \Gamma e B}{\pi T^3}
e^{-\frac{\sqrt{e B}}{T}}.
\end{equation}

\subsubsection{Numerical calculation of thermal conductivity}
\label{sec:kappa.numerical}

\begin{figure}[h]
\centering{
\includegraphics[width=8cm]{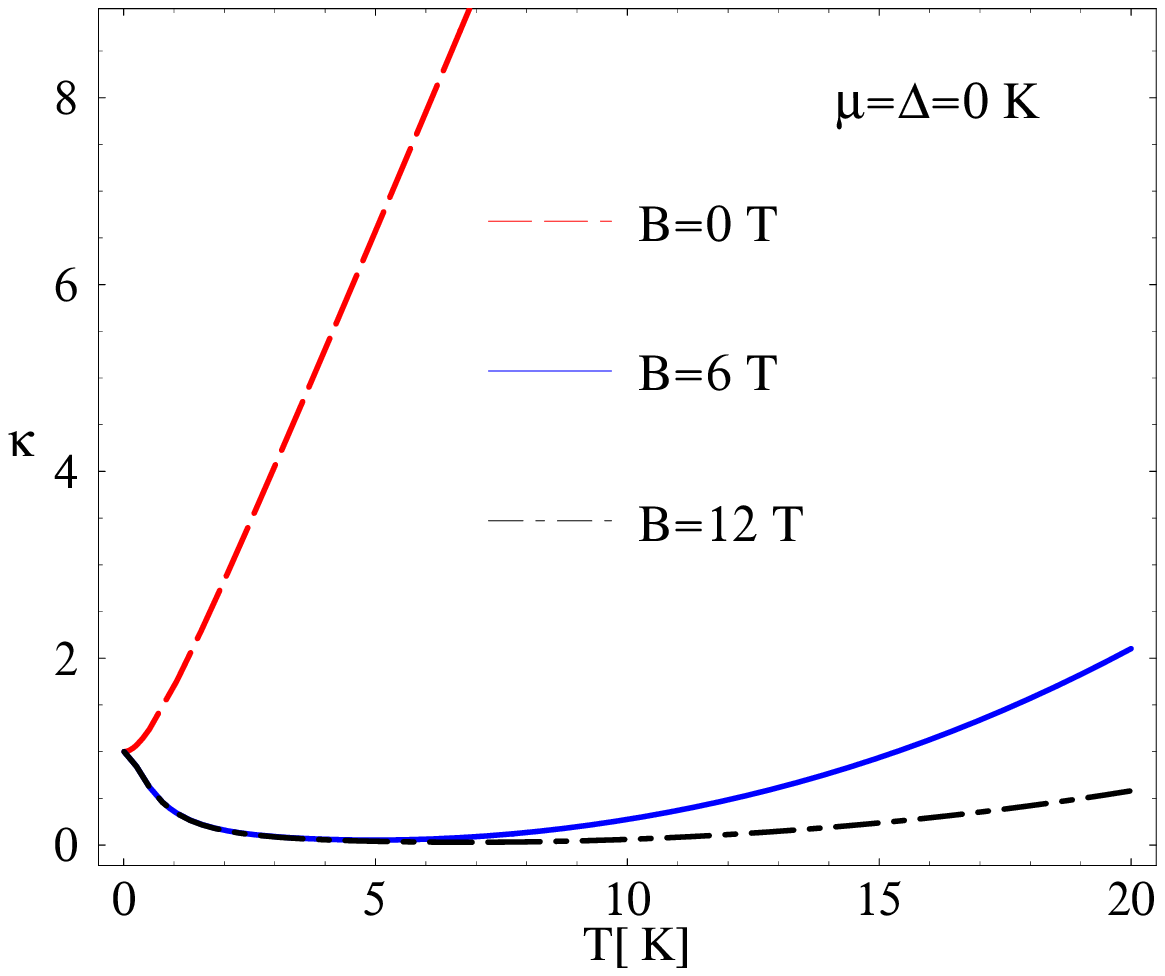}}
\caption{The normalized thermal conductivity $\kappa/\kappa(\mu, B =0, T \to 0)$
as function of temperature, $T$, for three different values of magnetic
field $B$ at half-filling, $\mu= 0$.
The value of $\kappa(\mu , B =0, T \to 0)$ is given by Eq.~(\ref{kappa.mu=0}).}
\label{fig:7}
\end{figure}

\begin{figure}[h]
\centering{
\includegraphics[width=8cm]{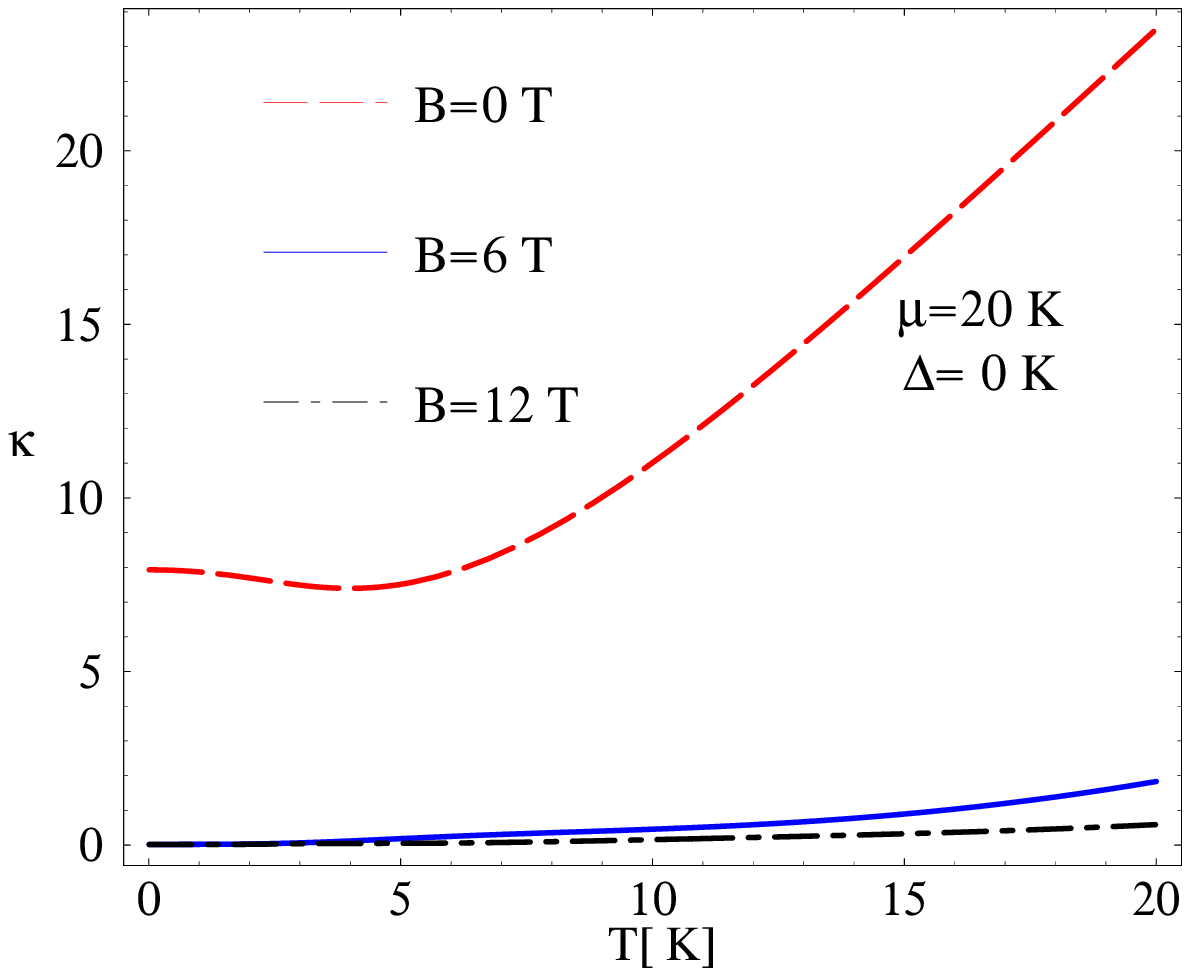}}
\caption{The normalized thermal conductivity $\kappa/\kappa(\mu, B =0, T \to 0)$
as function of temperature, $T$, for three different values of magnetic
field $B$ away from half-filling, $\mu= 20 \mbox{K}$.}
\label{fig:8}
\end{figure}

In Figs.~\ref{fig:7} and \ref{fig:8} we present the results for
the thermal conductivity that were obtained for the same values of
the model parameters as the data for electrical conductivity shown
in Figs.~\ref{fig:3} and \ref{fig:4}. Comparing these figures, one
can see that the behavior of the thermal conductivity is rather
similar to the behavior of electrical conductivity and to see more
subtle differences we have to consider the temperature dependence
of the Lorenz number as done in Sec.~\ref{sec:WF}.

\begin{figure}[h]
\centering{
\includegraphics[width=8cm]{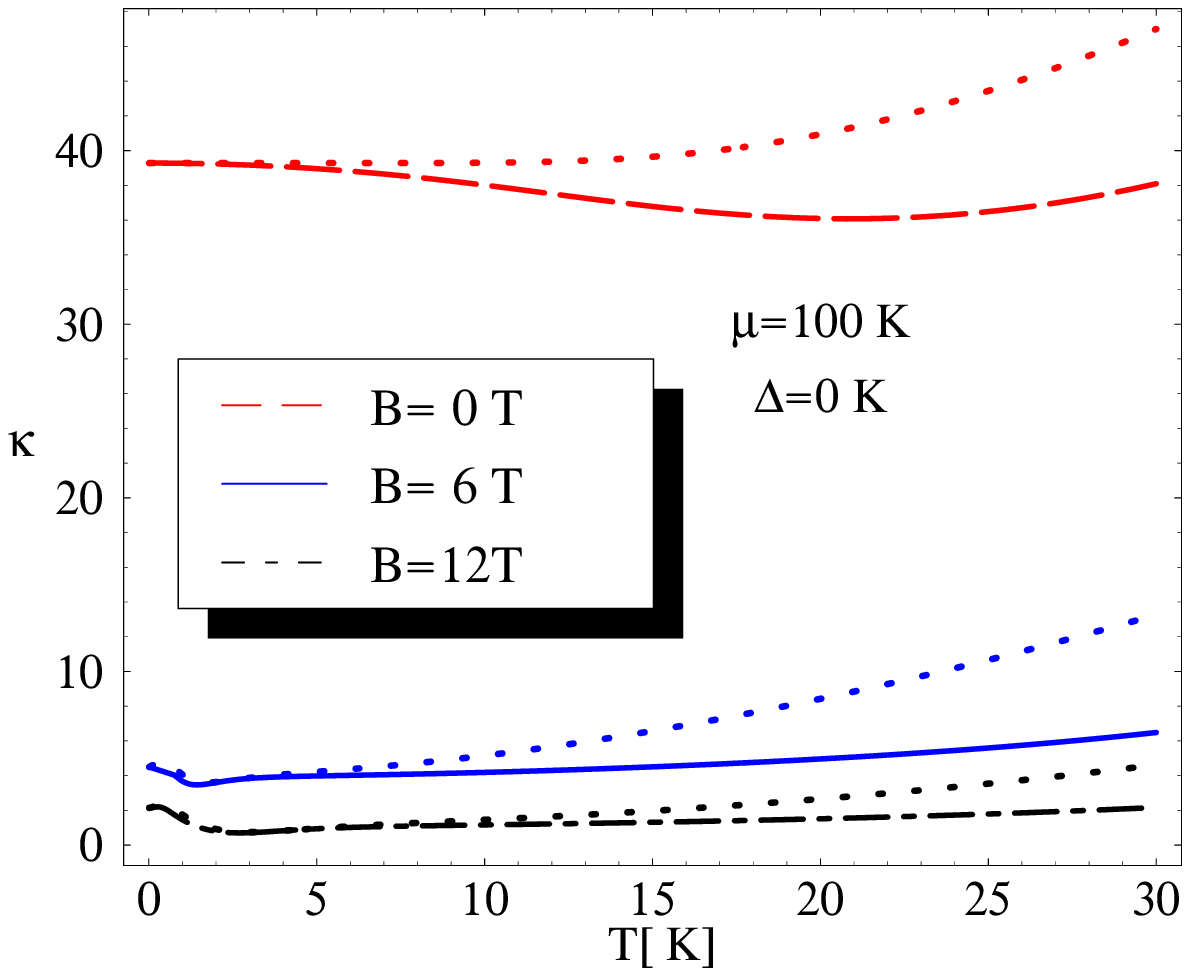}}
\caption{The normalized thermal conductivity $\kappa/\kappa(\mu, B =0, T \to 0)$
as function of temperature, $T$, for three different values of magnetic
field $B$ away from half-filling, $\mu= 100 \mbox{K}$.
The dotted lines are calculated without the second term of
Eq.~(\ref{thermal.conductivity.general.final3}) that originates from
the condition of absence of the  electrical current in the system.}
\label{fig:9}
\end{figure}

It is also interesting to investigate numerically the contribution
from the second term of
Eq.~(\ref{thermal.conductivity.general.final3}) to the thermal
conductivity. This is done in Fig.~\ref{fig:9}, where one can
clearly see that when the temperature grows and $T$ becomes
$\lesssim |\mu|$ the second term  of
Eq.~(\ref{thermal.conductivity.general.final3}) makes an important
negative contribution to the thermal conductivity.

\begin{figure}
\centering{
\includegraphics[width=8cm]{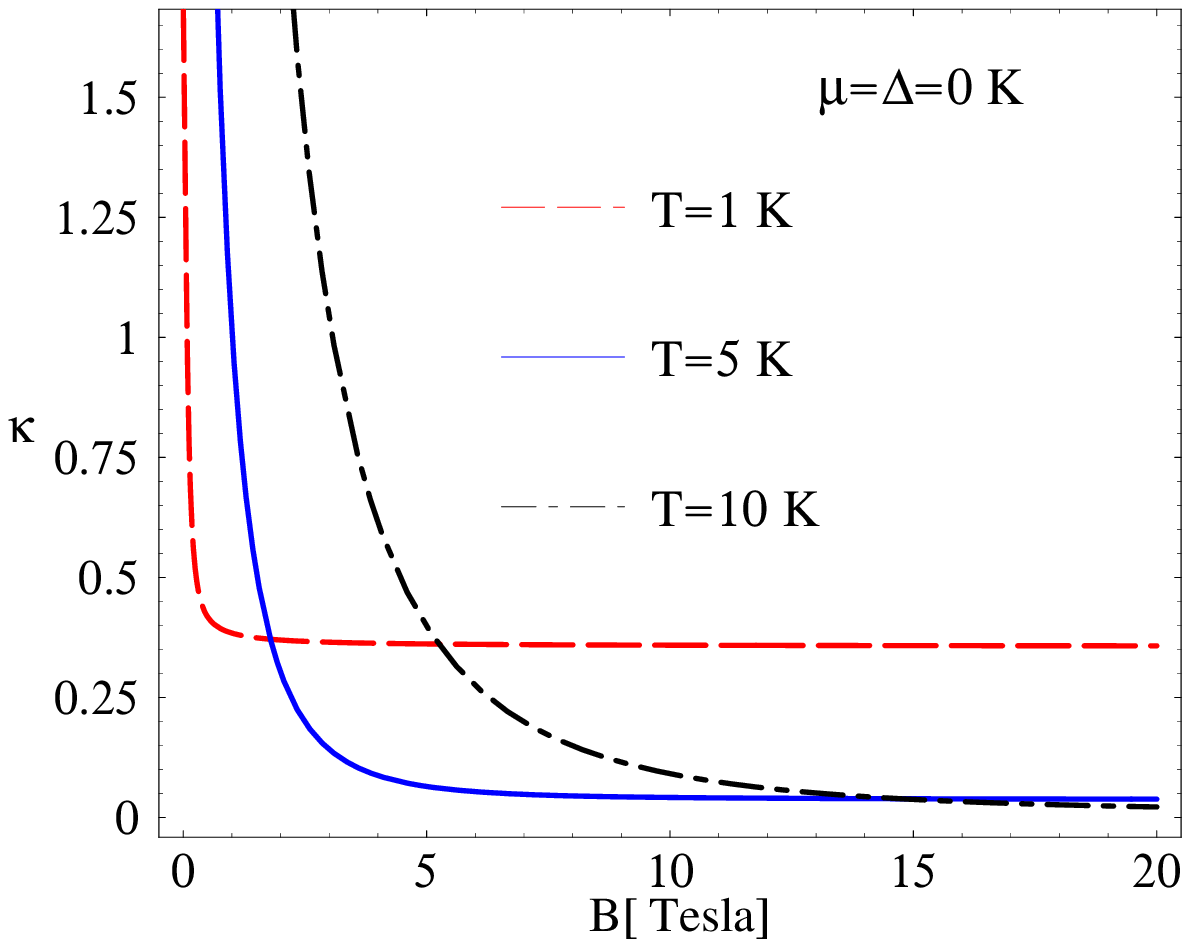}}
\caption{The normalized thermal conductivity $\kappa/\kappa(\mu, B
=0, T \to 0)$ as function of field, $B$, for three different values
of temperature $T$ at half-filling, $\mu= 0$.}
\label{fig:10}
\end{figure}

\begin{figure}[h]
\centering{
\includegraphics[width=8cm]{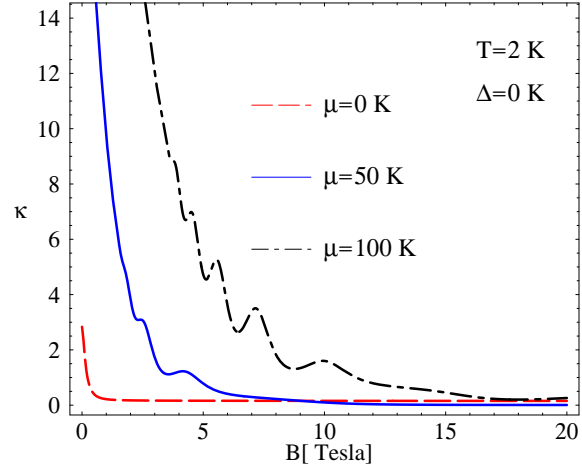}}
\caption{The normalized thermal conductivity $\kappa/\kappa(\mu, B
=0, T \to 0)$ as function of field, $B$, for three different values
of chemical potential $\mu$.}
\label{fig:11}
\end{figure}

In Figs.~\ref{fig:10} and \ref{fig:11} we present the results for the dependence
of thermal conductivity  on the magnetic field $B$
that were obtained for the same values of the model
parameters as the data for electrical conductivity
shown in Figs.~\ref{fig:5} and \ref{fig:6}.

\section{Wiedemann-Franz law}
\label{sec:WF}

After deriving in Secs.~\ref{sec:electrical} and \ref{sec:thermal}
the expressions for electrical and thermal conductivities in the
various limits we are ready to consider their implications for the
WF law.

\subsection{Limit $T \to 0$, arbitrary field, and chemical potential}

\begin{figure}[h]
\centering{
\includegraphics[width=8cm]{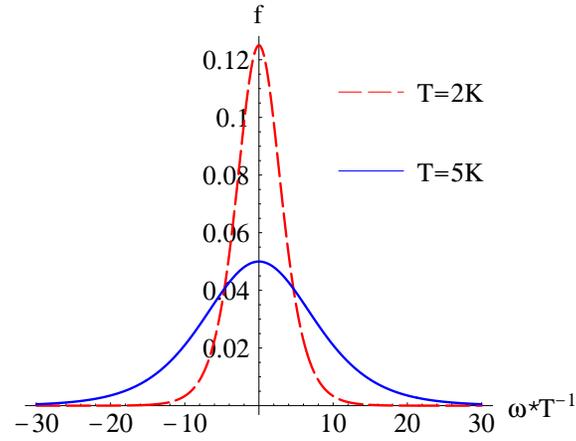}}
\caption{The expression for electrical conductivity contains the
factor $f(\omega) = -n_{F}^{\prime}(\omega-\mu)$ (the case $\mu
=0$ is plotted) and thus samples $\mathcal{A}(\omega)$ in the
immediate vicinity of the Fermi surface.} \label{fig:12}
\end{figure}

\begin{figure}[h]
\centering{
\includegraphics[width=8cm]{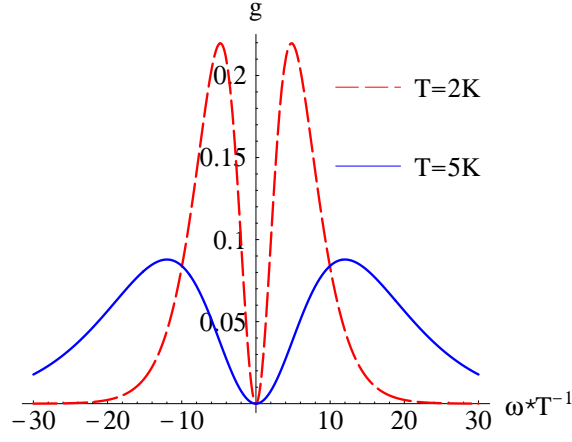}}
\caption{The expression for thermal conductivity contains the
factor $g(\omega) = -n_{F}^{\prime}(\omega-\mu)(\omega
-\mu)^2/T^2$ (the case $\mu =0$ is plotted), and thus measures
$\mathcal{A}(\omega)$ in the immediately below and above the Fermi
energy.} \label{fig:13}
\end{figure}

\begin{figure}[h]
\centering{
\includegraphics[width=8cm]{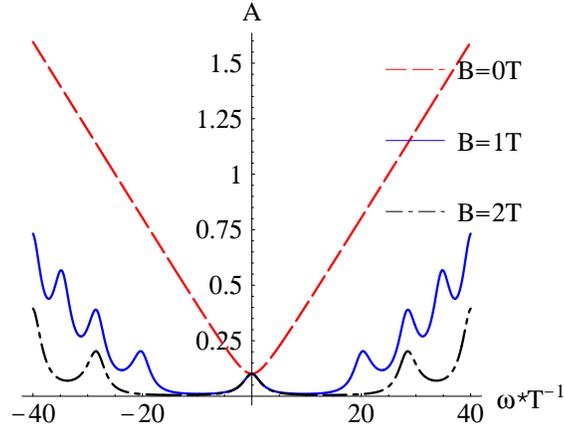}}
\caption{If $\mathcal{A}(\omega)$ varies appreciably over the
energy range from $\mu - k_B T$ to $\mu+ k_B T$ then the Lorenz
number deviates from its usual value $L_0$.} \label{fig:14}
\end{figure}

For finite $\Gamma$, arbitrary $\mu$, $B$, and $\Delta$ in the
limit $T \to 0$ one can see from Eq.~(\ref{electric.B.final}) [see
also Eq.~ (\ref{electric.B=0.final}) for $B =0$] and
(\ref{kappa.B.final}) [see also Eq.~(\ref{kappa.B=0.final}) for
$B=0$] that WF law is maintained:
\begin{equation}
\label{L0}
L_0 = \frac{\kappa}{\sigma T} = \frac{\pi^2}{3} \frac{k_B^2}{e^2} .
\end{equation}
This result can be understood from the qualitative arguments
given, for example, in Ref.~\cite{Taylor.book}. The expressions
for the electrical, Eq.~(\ref{sigma.final2}), and thermal,
Eq.~(\ref{thermal.conductivity.general.final3}), conductivities
calculated within the bare bubble approximation are very similar
in the limit $T \to 0$ when the second term of
Eq.~(\ref{thermal.conductivity.general.final3}) vanishes. Both
expressions contain the function $\mathcal{A}(\omega)$ multiplied
by the derivative of the Fermi distribution $f(\omega)=
-n_F^{\prime}(\omega -\mu)$, with the only difference that the
thermal conductivity is also multiplied by the factor $\sim
(\omega -\mu)^2$. These prefactors
 $f(\omega)$ and $g(\omega) = - (\omega - \mu)^2
n_F^{\prime}(\omega -\mu)/T^2$ are shown in Figs.~\ref{fig:12} and \ref{fig:13}.
When the temperature $T$ goes to zero, the $\delta$-like spikes of $f(\omega)$
and $g(\omega)$ occur at the same value $\omega = \mu$ and the WF law is maintained.

\subsection{Zero magnetic field at $T \neq 0$}

For $\mu = 0$ and $T \gg \Gamma$ considering
Eqs.~(\ref{kappa.T>Gamma.mu=0}) and (\ref{sigma.T>Gamma}) we
obtain that the WF law  is violated,
\begin{equation}
\label{Lorenz.mu=0}
L = \frac{\kappa}{\sigma T} = \frac{9 \zeta(3)}{2\ln 2}
\frac{k_B^2}{e^2} \simeq 2.37 L_0,
\end{equation}
while for $\mu \gg T$ Eqs.~(\ref{kappa.T>Gamma}) and (\ref{sigma.T>Gamma})
show that the Lorenz number has its usual value
\begin{equation}
\label{Lorenz.mu>T}
L = L_0.
\end{equation}
These two results can also be understood qualitatively using
Figs.~\ref{fig:12} and \ref{fig:13}. For $|\mu| \gg  T >0$ the
electrical conductivity is determined by energies $\omega \approx
\mu =0$, while the thermal conductivity is determined by the
energies near $\omega \approx \mu \pm k_B T$. If the function
$\mathcal{A}(\omega)$ does not vary appreciably over the energy
range $\mu - k_B T$ to $\mu + k_B T$, the Lorenz number may still
be $\sim 1$ even for finite $T$, $T \ll \mu$, as follows from
Eq.~(\ref{Lorenz.mu>T}). If $\mu$ becomes of the same order of
$T$, the dimensionless values $\mu/ T$ and $\mu/T \pm 1$ are apart
and the Lorenz number deviates from 1 as seen from
Eq.~(\ref{Lorenz.mu=0}).

For nonzero $\Delta \gg T$ and $\mu = 0$ we get from
Eqs.~(\ref{sigma.Delta}) and (\ref{kappa.Delta}) that
\begin{equation}
\label{Lorenz.Delta.mu=0}
L  = \frac{k_B^2}{e^2} \frac{\Delta^2}{T^2},
\end{equation}
i.e., $L > L_0$ for $\Delta > (\pi/\sqrt{3}) T$. This dominance of
the thermal conductivity can be easily understood  from the fact
that when the gap opens, the electrical conductivity  diminishes
more strongly because it is determined by the energies $\omega
\approx \mu$ near the Fermi surface.

\subsection{Nonzero magnetic field at $T \neq 0$}

For $\mu = \Delta =0$  in the {\it strong field limit}, $\sqrt{e
B} \gtrsim 4 T$ from Eqs.~(\ref{sigma.Bstrong}) and
(\ref{kappa.Bstrong}) we obtain that the Lorenz number becomes
field dependent:
\begin{equation}
L = \frac{32 eB}{T^2} e^{-\frac{\sqrt{e B}}{T}}.
\end{equation}
In the presence of magnetic field the function
$\mathcal{A}(\omega)$ (see Fig.~\ref{fig:14}) varies appreciably
over the energy range $\omega - k_B T$ to $\omega + k_B T$, so
that it is nonzero for $\omega = 0$ and very small at $\omega =
\pm k_B T$. This feature of $\mathcal{A}(\omega)$, along with the
small values of $\mu$, can produce rather strong violation of the
WF law.

\begin{figure}[h]
\centering{
\includegraphics[width=8cm]{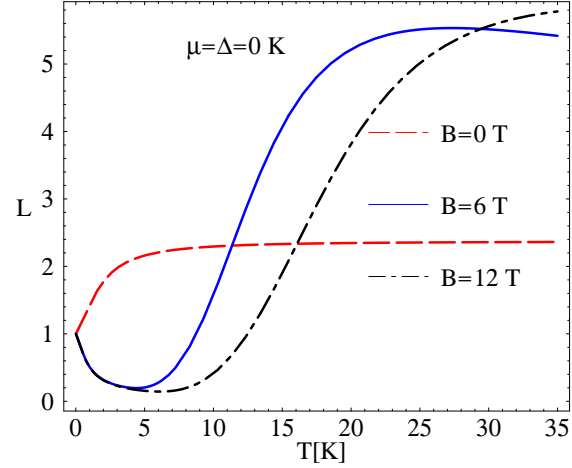}}
\caption{The normalized Lorenz number $L/L_0$
as function of temperature, $T$, for three different values of magnetic
field $B$ at half-filling, $\mu =0$.}
\label{fig:15}
\end{figure}

\begin{figure}[h]
\centering{
\includegraphics[width=8cm]{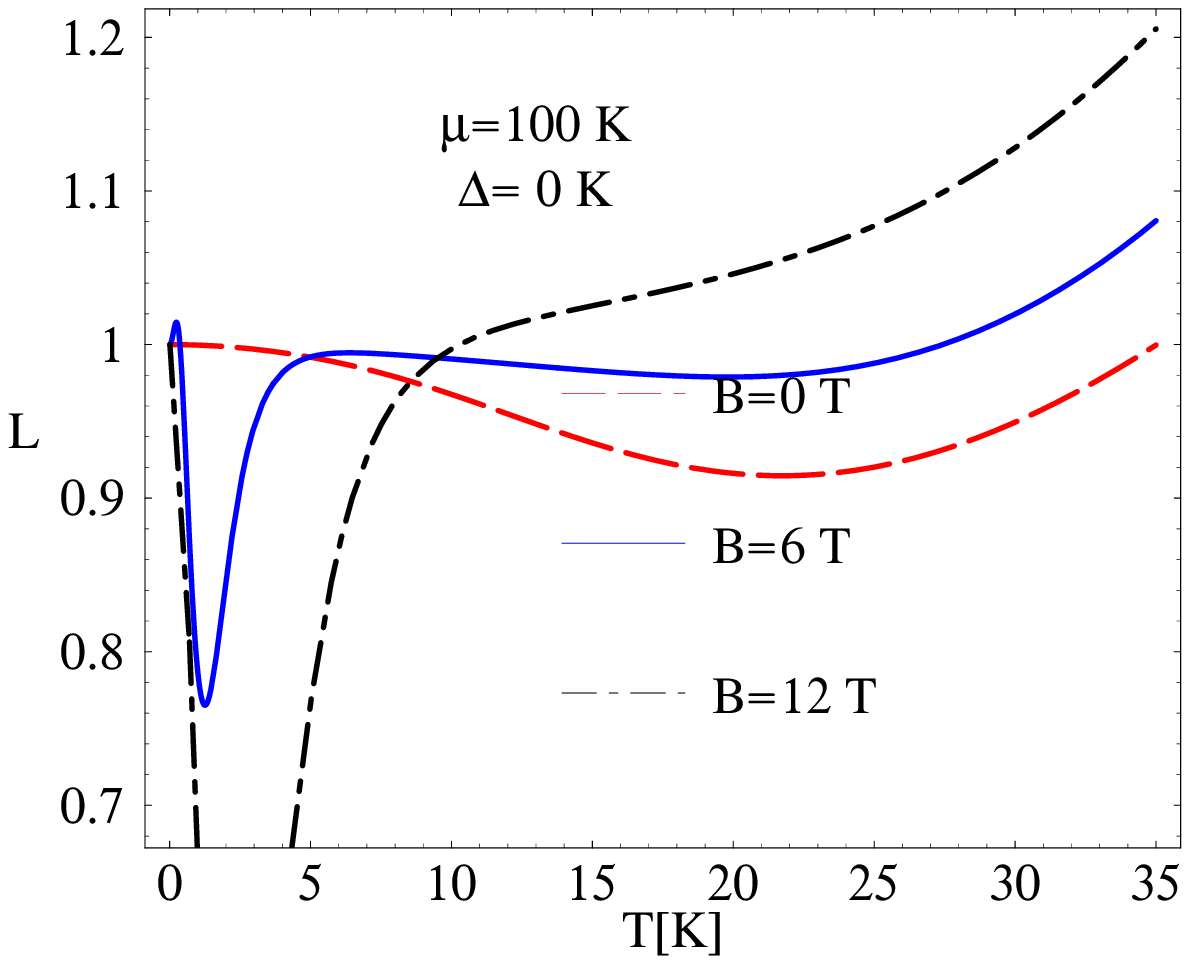}}
\caption{The normalized Lorenz number $L/L_0$
as function of temperature, $T$, for three different values of magnetic
field $B$ away from half-filling, $\mu = 100 \mbox{K}$.}
\label{fig:16}
\end{figure}

\begin{figure}[h]
\centering{
\includegraphics[width=8cm]{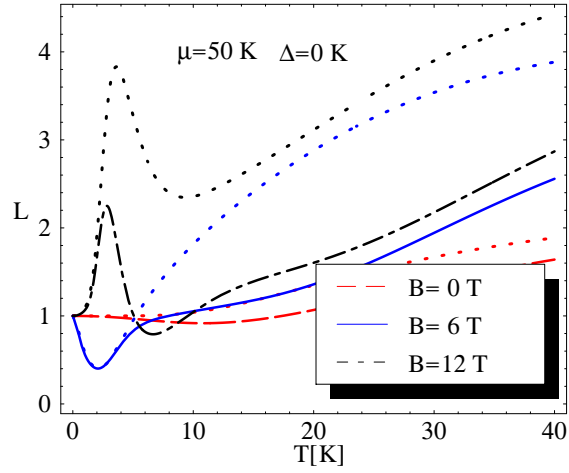}}
\caption{The normalized Lorenz number $L/L_0$
as function of temperature, $T$, for three different values of magnetic
field $B$ away from half-filling, $\mu= 50 \mbox{K}$.
The dotted lines are calculated without the second term of
Eq.~(\ref{thermal.conductivity.general.final3}) that originates from
the condition of absence of the  electrical current in the system.}
\label{fig:17}
\end{figure}

Our qualitative arguments are indeed confirmed by
Figs.~\ref{fig:15} and \ref{fig:16} where we show the temperature
dependence of $L(T)$ at half-filling (Fig.~\ref{fig:15}) and away
from it (Fig.~\ref{fig:16}). Both these figures are computed on
the basis of Eqs.~(\ref{sigma.final2}),
(\ref{thermal.conductivity.general.final3}), and (\ref{A.def}).
For $\mu =0$ the value of $L$ always goes down from $L_0$ as the
temperature increases, then $L$ increases crossing $L_0$, and
finally it goes to its zero-field value (\ref{Lorenz.mu=0}). One
can see from Fig.~\ref{fig:16} that for nonzero $\mu$ the
deviations of $L(T)$ from $L_0$ become less pronounced (see also
the discussion of Fig.~\ref{fig:17}).

It is also instructive to investigate the importance of the second term
of thermal conductivity, Eq.~(\ref{thermal.conductivity.general.final3}), for the
Lorenz number calculated with and without this term (as already done in
Fig.~\ref{fig:9} for the thermal conductivity itself). These results are given in
Fig.~\ref{fig:17} and they show that the second term of thermal conductivity is
indeed crucial to get the correct answer.
Note that for the case of nonzero $\mu$ the situation with the WF law
is more complicated because
the value of $\mu$ may coincide with $\omega$, corresponding to
the maximum of $\mathcal{A}(\omega)$. In this case $L(T)$ will
firstly increase, as one can see in Fig.~\ref{fig:17} for $B= 12 \mbox{T}$.

\section{Conclusions}
\label{sec:conclusions}

Let us compare the results derived in the present paper with the
experimental results obtained in Ref.~\cite{Hill:2001:Nature}. It
is still unknown whether the DDW state exists and/or plays an
important role in the electron doped compound used in
Ref.~\cite{Hill:2001:Nature}, so that making this comparison would
help to address these questions. First, we observe that while the
results of \cite{Hill:2001:Nature} suggest that the WF law is
violated at $T \to 0$, {\it there is no} violation of the WF law
in this limit in the DDW scenario of the pseudogap.   Since in
Fig.~3 of Ref.~\cite{Hill:2001:Nature} the electrical conductivity
is a constant, the line $\kappa_e(T)/T$ directly represents the
normalized Lorenz number $L(T)/L_0$. For finite temperatures there
is then some similarity between Fig.~\ref{fig:15} (or
Fig.~\ref{fig:16} for  $\mu \neq 0$ case) and Fig.~3 of
Ref.~\cite{Hill:2001:Nature} where as $T$ increases the thermal
conductivity crosses from the region with $\kappa_e/T <
L_0/\rho_0$ to the region with $\kappa_e(T)/T > L_0/\rho_0 $
resembling the character of the WF law violation seen in the
experiment.

However, in the experiment, the electrical conductivity is flat
while the thermal conductivity changes significantly in a
subkelvin range. On the other hand, theoretical calculations show
that the electrical (see Figs.~\ref{fig:3} and \ref{fig:4}) and
thermal (see Figs.~\ref{fig:7} and \ref{fig:8}) conductivities
vary simultaneously in a range of 10 K. Note that the wider range
of temperatures can be probably attributed to the fact that the
values of the model parameters we took for the numerical estimates
are more appropriate for the hole-doped compounds.

It is obvious from Fig.~\ref{fig:15} that such a behavior of
$L(T)$ is due to the {\it presence} of the magnetic field. This
confirms our claim that to interpret theoretically the experiment
in Ref.~\cite{Hill:2001:Nature} one should take into account the
influence of the external field.

Our main results can be summarized as follows.

\noindent (1) We have obtained analytical expressions for
electrical conductivity (\ref{sigma.final2}) and thermal
conductivity (\ref{thermal.conductivity.general.final3}) in the
DDW state in presence of an external magnetic field.

\noindent
(2) We have established a correspondence between the expression
for the electrical conductivity (\ref{sigma.like.Langer}) in the DDW state written
in terms of the generalized velocity $\mathbf{V}(\mathbf{k})$, Eq.~(\ref{velocity.general}),
and  the dc conductivity derived by Langer \cite{Langer:1962a:PR} in
the bubble approximation with the vertex (\ref{Lambda.Langer}).

\noindent (3) We have shown that in the DDW system in the presence
of impurities the WF law holds in $T \to 0$ limit for an arbitrary
field $B$ and chemical potential $\mu$. This is checked within the
bubble approximation, i.e., not including the impurity vertex. The
influence of the impurity vertex on the properties of the DDW
state in zero field at half-filling can be considered using the
heuristical arguments of Durst and Lee \cite{Durst:2000:PRB} for
$d$SC state. Since in the $d$SC state the thermal (and spin)
currents are proportional to the group velocity, they can relax
through either intranode scattering or scattering between nodes.
As a result, the different types of scattering play nearly the
same role and therefore vertex corrections do not contribute to
the thermal (and spin) conductivity. This is not the case of the
electrical current in the $d$SC state, because it depends only on
the Fermi velocity. Thus, the electrical current can relax more
effectively via scattering from node to node than it can via
scattering within a single node. This difference is taken into
account by considering vertex corrections which modify the bare
bubble expression for the electrical conductivity. Since the
electrical current in the DDW state is also proportional to the
group velocity, the corresponding impurity vertex should not
contribute to the electrical conductivity. Moreover, the thermal
current in the DDW state is also proportional to the group
velocity, so that one can rely on the same arguments about the
impurity vertex as for the $d$SC case. Thus we do not expect that
in the DDW state the impurity vertex can be a source of the WF
violation at $T \to 0$. These observations should be confirmed by
detailed calculations of the impurity vertex corrections, which,
however, are beyond the scope of the present paper.

\noindent (4) For finite temperatures $T \lesssim |\mu|$, the WF
law violation is possible and in zero field the thermal
conductivity dominates over the electrical conductivity, i.e.,
$L(T)/L_0 > 1$.

\noindent
(5) For  $T \lesssim |\mu|$
in the nonzero field the WF law violation becomes even
stronger than in zero field and depending on the temperature both regimes
 $L(T)/L_0 \ll 1$  and $L(T)/L_0 > 1$ are possible.

\noindent
(6) For $T \ll |\mu|$ there is no WF violation even in the presence of
magnetic field.

Finally, we would like to stress that the results of the present
paper should be applicable not only to the DDW state, but to a
wider class of  theories. Here we started from the DDW Hamiltonian
(\ref{Hamiltonian.DDW}) and, to simplify the problem, approximated
it by the QED$_3$ Lagrangian (\ref{Dirac.conjugated.Lagrangian}).
There is, however, a number of systems, e.g., pyrolitic graphite
\cite{Gorbar:2002:PRB}, that can also be described by the
Lagrangian  (\ref{Dirac.conjugated.Lagrangian}), so that our
results would also be relevant for them.

\section{Acknowledgments}
We gratefully acknowledge A.~Gr\"oger for a stimulating
discussion. S.G.Sh would like to thank N.~Andrenacci, L.~Benfatto
and L.~Carlevaro for helpful discussions and W.~Kim for useful
correspondence. This work was supported by  Research Project No.
20-65045.01 of the Swiss NSF. The work of V.P.G. was supported by
the SCOPES Projects No. 7UKPJ062150.00/1 and No. 7 IP 062607 of
the Swiss NSF and by Grant No. PHY-0070986 of NSF (USA).

\appendix
\section{Nodal fermion Green's function in an external magnetic field}
\label{sec:A}

In the coordinate space, the fermion propagator has the following
form in the proper-time representation
\begin{equation}
\label{Schwinger.1}
S(x,y) = (i \hat{D} + \Delta)_x
\langle x\left|\frac{-1}{\Delta^2 + \hat{D}^2} \right|y\rangle =
-i (i \hat{D} + \Delta)_x
\int \limits_{0}^{\infty} d s \langle x| \exp [ -is(\Delta^2 + \hat{D}^2)] |y\rangle,
\end{equation}
where $\hat{D} = \gamma^{\nu} D_{\nu}$ and the `'long'' derivative
$D_{\nu}$ is given by Eq.~(\ref{long.derivative}). The matrix
element $\langle x| \exp [ -is(\Delta^2 + \hat{D}^2)] |y\rangle$
can be calculated using the Schwinger (proper time) approach
\cite{Schwinger:1951:PR} (see also a pedagogical overview in
Ref.~\cite{Dittrich.book}). The main idea of this method is based
on the interpretation of $\langle x | \exp [ -is \hat{D}^2]
|y\rangle$ as the coordinate representation of the proper-time
``evolution operator'' $U(s) = \exp[- i Hs]$, where we have
introduced the ``Hamiltonian''
\begin{equation}
H = \hat{D}^2 =  D^2 - \frac{e}{2} \sigma^{\rho \nu} F_{\rho \nu},
\qquad F_{\rho \nu} = \partial_{\rho} A_{\nu} - \partial_{\nu} A_{\rho},
\qquad \sigma^{\rho \nu} = \frac{i}{2} [\gamma^{\rho}, \gamma^{\nu}].
\end{equation}
The matrix element of the evolution
operator can be evaluated \cite{Dittrich.book}, for example,
using either operator or functional integral formalism:
\begin{equation}
\label{U.final}
\langle x | U(s) | y \rangle = \frac{e^{-i \pi/4}}{8 (\pi s)^{3/2}}
\exp \left[ i e \int \limits_y^x d \xi_{\nu} A^{\nu} (\xi) +
\frac{i}{4}(x-y) f(s) (x-y) - L(s) + i \frac{e}{2}
\mathbb{\sigma} \mathbb{F} s \right],
\end{equation}
with
\begin{equation}
f(s) \equiv e \mathbb{F} \coth (e \mathbb{F} s), \qquad
L(s) \equiv \frac{1}{2} \mbox{tr} \ln
\left(\frac{\sinh e \mathbb{F}s}{e \mathbb{F} s}\right),
\end{equation}
where we have used matrix notations, e.g., $F_{\rho \nu} \equiv
(\mathbb{F})_{\rho\nu}$ and the integral with $A_{\nu}$ is
calculated along the straight line. This operator differs from the
four-dimensional version of $\langle x | U(s) | y \rangle$
\cite{Dittrich.book} in the power dependence of the proper time
and in the numerical prefactors.
 Plugging (\ref{U.final}) into (\ref{Schwinger.1}) one obtains
\begin{equation}
\label{Schwiner.2}
\begin{split}
S(x,y) &= \exp \left( i e \int \limits_y^x d \xi_{\nu} A^{\nu} (\xi) \right)
\frac{e^{-i3 \pi/4}}{8 \pi^{3/2}} \int \limits_{0}^{\infty} \frac{ds}{s^{3/2}}
\left[ \Delta - \frac{1}{2} \gamma^{\rho} [f(s) + e \mathbb{F}]_{\rho \nu}
(x-y)^{\nu}\right] \\
& \times \exp \left[- i \Delta^2 s
+ \frac{i}{4}(x-y) f(s) (x-y) - L(s) + i \frac{e}{2}
\mathbb{\sigma} \mathbb{F} s \right].
\end{split}
\end{equation}

Finally, to perform calculations for the case of interest one should
evaluate $f(s)$, $L(s)$ and $\exp [i (e/2) \mathbb{\sigma} \mathbb{F} s]$
for $\mathbb{F}$ containing purely magnetic background field $B$.
The corresponding expressions  for these values are
\begin{equation}
\begin{split}
& f(s)_{\rho \nu} = - \frac{1}{s} \left[ g_{\rho \nu} +
\frac{(\mathbb{F}^2)_{\rho \nu}}{B^2} (1 - eBs \cot (eBs))
\right],\\
& \exp[-L(s)] = \frac{e B s}{\sin e Bs}, \\
& \exp \left[i \frac{e}{2} \mathbb{\sigma} \mathbb{F} s \right] =
\cos e B s + \gamma^1 \gamma^2 \sin e B s.
\end{split}
\end{equation}
In the Matsubara frequency-momentum representation this leads to
Eq.~(\ref{Schwinger.representation.translation}).

It is convenient to write down an alterative form of the fermion
propagator in a magnetic field as a sum over the Landau-level
poles. In deriving it we follow
Refs.~\cite{Gusynin:1995:PRD,Chodos:1990:PRD}. Introducing the
shorthand notations we rewrite
Eq.~(\ref{Schwinger.representation.translation})  as follows,
\begin{equation}
\begin{split}
{\tilde S}(i \omega, \mathbf{p})
= & -  \int \limits_{0}^{\infty} d s
\exp\left[- s\left(a + \mathbf{p}^2 \frac{\tanh(eBs)}{eBs}\right) \right]
\times \left[ \left( b + c  \tanh(|eBs|) \right)
\left(1 + d \tanh (|eBs|) \right) \right],
\end{split}
\end{equation}
where
\begin{equation}
\begin{split}
& a = \Delta^2 - (i\omega)^2, \qquad
b =  i \omega \gamma^0 - p_1 \gamma^1 - p_2 \gamma^2 + \Delta,\\
& c = -i(p_2 \gamma^1- p_1 \gamma^2) \mbox{sgn}(e B), \qquad d = -i\gamma^1 \gamma^2
\mbox{sgn}(e B).
\end{split}
\end{equation}
Using the identity $\tanh x = 1 - 2 \exp(-2x) /[1 + \exp(-2x)]$, the relation
\cite{Gradshtein.book}
\begin{equation}
(1-z)^{- (\alpha+1)} \exp \left(\frac{xz}{z-1}\right) = \sum_{n=0}^{\infty}
L_n^{\alpha} (x) z^n, \qquad |z| < 1,
\end{equation}
with $z = - \exp(-2|eB|s)$, $x = 2 \frac{\mathbf{p}^2}{|eB|}$, we arrive at
the expression
\begin{equation}
\begin{split}
&{\tilde S}(i \omega, \mathbf{p}) = - \exp\left( - \frac{\mathbf{p}^2}{|e B|}\right) \\
& \times \int \limits_0^{\infty} d s \exp\left[ -s a + \frac{2\mathbf{p}^2}{|e B|}
\frac{z}{z-1} \right]
\left[ \left(b + c\right) \left( 1 + d \right) +
\left(4cd  +2c + 2bd \right) \frac{z}{1-z} +
4 cd \frac{z^2}{(1-z)^2}\right].
\end{split}
\end{equation}
Then using the identity $L_n^{\alpha-1}(x) = L_n^{\alpha}(x) - L_{n-1}^{\alpha}(x)$
and the definitions $L_n \equiv L_n^0$, $L_{-1}^{\alpha} =0$
we obtain
\begin{equation}
\begin{split}
{\tilde S} & (i \omega, \mathbf{p}) = -\exp\left( - \frac{\mathbf{p}^2}{|e B|}\right)
\int \limits_{0}^{\infty} d s \exp (-sa)
\left[ \left(b + c\right) \left( 1 + d \right) \sum_{n=0}^{\infty}
L_n(x) z^n \right. \\
& \left. + \left(-\left(b + c\right) \left( 1 + d \right)
  + 2c + 2bd \right) \sum_{n=0}^{\infty} L_{n-1}(x)z^n
+ 4cd \sum_{n=0}^{\infty} L_{n-1}^1 (x) z^n \right] \\
 = &  - \exp\left( - \frac{\mathbf{p}^2}{|e B|}\right)
\int \limits_{0}^{\infty} d s \exp (-sa)
\left\{ (\Delta + i \omega \gamma^0) \left[ (1- \mbox{sgn}(e B) i \gamma^1 \gamma^2)
\sum_{n=0}^{\infty} L_n(x) z^n - (1+  \mbox{sgn}(e B) i \gamma^1 \gamma^2)
\sum_{n=0}^{\infty} L_{n-1}(x)z^n
\right] \right. \\
& \left. + 4 (p_1 \gamma^1 + p_2 \gamma^2) \sum_{n=0}^{\infty} L_{n-1}^1 (x) z^n
\right\}.
\end{split}
\end{equation}
Finally, integrating over $s$ we arrive at Eq.~(\ref{Landau.levels}).

\section{Generalized polarization  bubble and nodal approximation}
\label{sec:B}

The calculations of  electrical, thermal, and even spin
conductivities are quite similar for both DDW and $d$SC \cite{Durst:2000:PRB}
nodal systems. Thus  instead of
repeating the same calculation several times and to underline the
similarities and differences between DDW and $d$SC cases, it is
rather convenient to define a generalized polarization tensor
$\Pi^{g g^{\prime}}$ that depends on the generalized coupling parameters
$g$, $g^{\prime}$
\begin{equation}
\label{couplings}
g , g^{\prime} = \left[e,
\begin{cases}
i \omega + i \Omega/2\\
\omega + \Omega/2 - \mu
\end{cases} \right].
\end{equation}
Two lines in the definition of $g, g^{\prime}$ correspond to the
Matsubara and real frequencies, respectively, and the origin of
the chemical potential $\mu$ in the second line will become clear
later. The generalized bubble is
\begin{equation}
\label{generalized.bubble}
\Pi^{g g^{\prime}}_{\alpha \beta}(i\Omega) = 2 T
\int_{\mathrm{RBZ}} \frac{d^2 k}{(2 \pi)^2}
\sum_{i \omega} g g^{\prime}
\mbox{tr} [G(i \omega, \mathbf{k}) V_{\alpha}(\mathbf{k})
G(i \omega + i \Omega, \mathbf{k}) V_{\beta}(\mathbf{k})],
\end{equation}
where $G(i\omega, \mathbf{k})$ is the Green's function
(\ref{Green.common}), but evaluated in the external field, and
$\mathbf{V}(\mathbf{k})$ is generalized velocity given by
Eq.~(\ref{velocity.general}). The integral is over the reduced
Brillouin zone (RBZ) and the factor 2 before the integral is due
to the spin degree of freedom $s$. To compare this expression with
$d$SC case [see Eq.~(A1) of Ref.~\cite{Durst:2000:PRB}] we note
that the summation over the spin degree of freedom {\it is already
included} in the Nambu formalism.

Using the spectral representation for the fermion Green's function
\begin{equation}
\label{G.translation.invariant}
G(i \omega + \mu, {\bf k}) =  \int \limits_{-\infty}^{\infty}
\frac{A(\omega_1, {\bf k})}{i\omega + \mu - \omega_{1}}d\omega_{1},
\end{equation}
where $A(\omega_1, {\bf k})$ is the spectral density,
we arrive at
\begin{equation}
\Pi^{g g^{\prime}}_{\alpha \beta}(i \Omega) = 2 \int_{\mathrm{RBZ}} \frac{d^2 k}{(2 \pi)^2}
\int d\omega_{1} \int d\omega_{2}
\mbox{tr} \left[ A(\omega_1, {\bf k}) V_{\alpha}(\mathbf{k})
A(\omega_2, {\bf k}) V_{\beta}(\mathbf{k})
\right] R
\end{equation}
with
\begin{equation}
\label{Matsubara.sum}
R = T \sum_{i\omega} g g^{\prime} \frac{1}{i\omega + \mu - \omega_{1}}
\frac{1}{i\omega + \mu + i \Omega  - \omega_{2}}.
\end{equation}

Since the intermediate results differ depending on the frequency
dependence of the coupling parameters $g$, $g^{\prime}$, we
consider the frequency-independent and frequency-dependent
couplings separately. For $g=g^{\prime} = e$ (frequency
independent coupling), evaluating the sum and then continuing $i
\Omega \to \Omega+ i 0$ we get
\begin{equation}
\label{R.independent}
R =
g^2 \frac{n_{F}(\omega_1 - \mu) - n_F(\omega_2 - \mu)}{\omega_1 - \omega_2 +
\Omega + i0} =
\frac{g^2}{2} \frac{\tanh \frac{\omega_2-\mu}{2T} - \tanh\frac{\omega_1- \mu}{2T}}
{\omega_1 - \omega_2 +  \Omega + i 0}.
\end{equation}
For the frequency-dependent coupling $g = g^{\prime}= i \omega + i
\Omega/2$ the evaluation of the Matsubara sum
(\ref{Matsubara.sum}) gives \cite{foot2}
\begin{equation}
R = T \sum_{i\omega} \frac{(i \omega + i \Omega/2)^2}
{(i\omega + \mu - \omega_{1})(i\omega + \mu + i \Omega  - \omega_{2})} =
\frac{(\omega_1 - \mu + i \Omega/2)^2 n_F(\omega_1 - \mu) -
(\omega_2 - \mu - i \Omega/2)^2 n_F(\omega_2 - \mu)}
{\omega_1 - \omega_2 + i \Omega}.
\end{equation}
Then, continuing to real frequencies $i \Omega \to \Omega + i0$ we obtain
\begin{equation}
\label{R.dependent}
\begin{split}
\mbox{Im} R
= \frac{\pi}{2} (\omega_1 - \mu + \Omega/2)^2 \left[
\tanh \frac{\omega_1 - \mu}{2T} - \tanh \frac{\omega_1 -\mu + \Omega}{2T}
\right] \delta (\omega_1 - \omega_2 + \Omega).
\end{split}
\end{equation}
Comparing the imaginary part of Eq.~(\ref{R.independent}) with Eq.~(\ref{R.dependent})
one can see that the latter equation can be obtained from the former
by the direct substitution of $g = g^{\prime} = \omega + \Omega/2 - \mu$.
This explains the difference between the upper and lower lines in Eq.~(\ref{couplings}).

Similarly, evaluating the Matsubara sum (\ref{Matsubara.sum}) with
$g = e$, $g^{\prime}= i \omega + i \Omega/2$
we get
\begin{equation}
R = T \sum_{i\omega} \frac{e  (i \omega + i \Omega/2)}
{(i\omega + \mu - \omega_{1})(i\omega + \mu + i \Omega  - \omega_{2})} =
\frac{e(\omega_1 - \mu + i \Omega/2) n_F(\omega_1 - \mu) -
e(\omega_2 - \mu - i \Omega/2) n_F(\omega_2 - \mu)}
{\omega_1 - \omega_2 + i \Omega},
\end{equation}
so that
\begin{equation}
\begin{split}
\mbox{Im} R  = \frac{\pi}{2} e(\omega_1 - \mu + \Omega/2) \left[
\tanh \frac{\omega_1 - \mu}{2T} - \tanh \frac{\omega_1 -\mu + \Omega}{2T}
\right] \delta (\omega_1 - \omega_2 + \Omega).
\end{split}
\end{equation}
Finally, we obtain for the imaginary part of the tensor polarization
$\Pi_{R}^{g g^{\prime}}(\Omega + i0)$ the following expression
\begin{equation}
\label{bubble.final.tensor}
\mbox{Im} \Pi^{g g^{\prime}}_{\alpha \beta}(\Omega + i 0) =
\pi \int_{\mathrm{RBZ}} \frac{d^2 k}{(2 \pi)^2}
\int \limits_{-\infty}^{\infty} d\omega g g^{\prime}
\left[ \tanh \frac{\omega - \mu}{2T} - \tanh \frac{\omega -\mu + \Omega}{2T} \right]
\mbox{tr} \left[ A(\omega, {\bf k}) V_{\alpha}(\mathbf{k})
A(\omega + \Omega, {\bf k}) V_{\beta}(\mathbf{k})  \right] .
\end{equation}

Looking at Eqs.~(\ref{generalized.bubble}) and
(\ref{bubble.final.tensor}) one can notice that the Green's
function $G$ and the associated spectral density $A$ in the
external field are in fact unknown. Instead of considering these
functions, we have constructed in Sec.~\ref{sec:Green.magnetic}
the Green's function (\ref{Landau.levels}) for the linearized
nodal Lagrangian (\ref{Dirac.conjugated.Lagrangian}) that is valid
in the vicinity of the four nodal points. Thus, in
Eq.~(\ref{generalized.bubble}) we replace the integration over the
reduced Brillouin zone by the integral over the $\mathbf{k}$-space
surrounding each node and sum over the four nodal subzones:
\begin{equation}
\label{scaling} \int_{\mathrm{RBZ}} \frac{d^2 k}{(2 \pi)^2}
\rightarrow \frac{1}{2}\sum_{j=1}^{4}\int\frac{d k_x d
k_x}{(2\pi)^{2}} \rightarrow
\frac{1}{2}\sum_{j=1}^{4}\int\frac{d^{2}p}{(2\pi)^{2}v_{F}v_{D}} =
\frac{1}{4 \pi v_F v_{D}} \sum_{j=1}^{4}\int \limits_0^{p_0} p dp
\int \limits_{0}^{2\pi} \frac{d \theta}{2 \pi},
\end{equation}
where $p_1 = v_{F} k_x = p \cos \theta$, $p_2 = v_{D} k_y = p \sin
\theta$, $p = \sqrt{p_1^2 + p_2^2}$, $p_0 = \sqrt{\pi v_{F}
v_{D}}/ (2 a)$ and the local nodal coordinate systems $(k_x, k_y)$
are shown in Fig.~\ref{fig:2}. Note again that, comparing with
\cite{Durst:2000:PRB}, an extra factor $1/2$  appears due to the
fact that the original integral is over the reduced Brillouin
zone.

The advantage of the scaled variables $p_1$ and $p_2$ is that they
correspond to the ``relativistic''  case $v_F = v_D = c =1$.
Moreover, we can use the spectral function $A_D$ in
Eq.~(\ref{AD.clean}) inside the integral over $p$, so that all
necessary factors with $v_{F}$ and $v_{D}$ are already outside the
integral (\ref{scaling}). We should only provide the rule that
allows us to restore the model parameters coming along with the
magnetic field in the final result \cite{Liu:1999:NP}: $e B \to
(\hbar v_F v_D/c)e B$.

We replace the generalized velocities $\mathbf{V}(\mathbf{k})$
in Eq.~(\ref{bubble.final.tensor})
by their
values $\mathbf{V}_{\mathbf{k} = \mathbf{N}}$
on the Fermi surface at half-filling ($\mu =0$).
Note that for the DDW case this approximation is more severe than
for the $d$SC case because for $\mu \neq 0$ the true Fermi surface does not coincide
with the Fermi surface at half-filling.
This approximation puts some restrictions on the values of $\mu$, so that
we cannot move far away from half-filling. The bubble (\ref{bubble.final.tensor})
takes the form
\begin{equation}
\label{bubble.nodal}
\mbox{Im} \Pi^{g g^{\prime}}_{\alpha \beta}(\Omega + i0) =  \frac{\pi }{2 v_{F}v_{D}}
\sum_{j=1}^{4}
\int \frac{d^{2}p}{(2\pi)^{2}} \int \limits_{-\infty}^{\infty} d\omega g g^{\prime}
\left[ \tanh \frac{\omega - \mu}{2T} - \tanh \frac{\omega -\mu + \Omega}{2T} \right]
\mbox{tr} \left[ A_D(\omega, {\bf p}) \gamma^{0} V_{\alpha}
A_D(\omega + \Omega, {\bf p}) \gamma^{0} V_{\beta}  \right] ,
\end{equation}
where the spectral density (\ref{AD.clean}) includes the
broadening of the spectral lines (\ref{Gamma}) due to impurities
and we inserted the $\gamma^0$ matrix [see Eq.~(\ref{G.B=0})].
Note that in Eq.~(\ref{G.translation.invariant}) we used only the
translationary invariant part
(\ref{Schwinger.representation.translation}) of the Green's
function (\ref{Schwinger.representation}) calculated in the
external field, since its translation non-invariant part cancels
out when  substituted in the bubble $\Pi$.

Since we are interested only in the longitudinal conductivities
and because the system is isotropic, we can define the
longitudinal polarization function $\Pi^{g g^{\prime}}$ as follows
(the sum over dummy index is implied):
\begin{equation}
\label{tensor2scalar}
\Pi^{g g^{\prime}}(\Omega) \equiv \frac{1}{2}
\Pi^{g g^{\prime}}_{\alpha \alpha}(\Omega).
\end{equation}
Then evaluating the sum over nodes in Eq.~(\ref{bubble.nodal}) by using
the identities
\begin{equation}
\label{nodal.sum}
\sum_{j=1}^{4} v_{l \alpha}^{(j)}  v_{l \beta}^{(j)}
= 2v_{l}^{2} \delta_{\alpha \beta}  , \quad
\sum_{j=1}^{4} v_{l \alpha}^{(j)}  v_{l^{\prime} \beta}^{(j)}
= 2 v_l v_{l^{\prime}} \epsilon_{l l^{\prime}} \epsilon_{\alpha \beta}
\quad (l \neq l^{\prime}),
\qquad \mathbf{v}_{l} \equiv \{\mathbf{v}_F, \mathbf{v}_D\},
\end{equation}
with $\epsilon_{ij}$ being antisymmetric tensor,
we arrive at the final result
\begin{equation}
\label{bubble.final}
\begin{split}
& \mbox{Im} \Pi^{g g^{\prime}}(\Omega + i0) =  \frac{\pi}{v_{F}v_{D}}
\int \frac{d^{2}p}{(2\pi)^{2}} \int \limits_{-\infty}^{\infty} d\omega g g^{\prime}
\left[ \tanh \frac{\omega - \mu}{2T} - \tanh \frac{\omega -\mu + \Omega}{2T} \right] \\
& \times \left( v_F^2 \mbox{tr} \left[ A_D(\omega, {\bf p}) \gamma^{1}
A_D(\omega + \Omega, {\bf p}) \gamma^{1}  \right] +
v_D^2 \mbox{tr} \left[ A_D(\omega, {\bf p}) \gamma^{2}
A_D(\omega + \Omega, {\bf p}) \gamma^{2}  \right] \right).
\end{split}
\end{equation}
Since both the electrical current (\ref{electric.current.DDW}) and
thermal current (\ref{heat.current2.B=0}) have two terms, a
``Fermi'' term proportional to $\mathbf{v}_F$ and $\sigma_3$ and a
``gap'' term proportional to $\mathbf{v}_D$ and $\sigma_2$, our
calculation of the current-current polarization function
(\ref{generalized.bubble}) is in fact similar to the calculation
of the thermal current bubble for the $d$SC case done in
Ref.~\cite{Durst:2000:PRB}. Evaluating the polarization function
(\ref{bubble.nodal}) we obtain four bubbles: Fermi-Fermi,
Fermi-gap, gap-Fermi, and gap-gap. However, since $\mathbf{v}_F
\perp \mathbf{v}_D$ at each of the gap nodes, the two cross terms
cancel,  as reflected in the second identity in
Eq.~(\ref{nodal.sum}). Therefore both electrical and thermal
longitudinal conductivities have two terms: the Fermi term with
the velocity $v^2_F$ and the gap term with the velocity $v^2_D$.
Thus, finally we approximated the polarization bubble for the
Hamiltonian (\ref{Hamiltonian.DDW}) by the polarization bubble for
the nodal Lagrangian (\ref{Dirac.conjugated.Lagrangian}).

\section{Some integrals}
\label{sec:C}

Here we derive the expression (\ref{kappa.T>Gamma}) for $\kappa$.
It follows from the general expression (\ref{thermal.conductivity.general.final3})
with $\mathcal{A}$ given by  Eq.~(\ref{A.B=D=0.T>Gamma}).
The corresponding integrals
are easily evaluated in terms of polylogarithmic functions, $\mbox{Li}_{n}(z)$,
\cite{Wolfram}:
\begin{equation}
\label{J}
\begin{split}
J & =\int \limits_{-\infty}^\infty\frac{d\omega(\omega-\mu)^2|\omega|}
{\cosh^2((\omega-\mu) /2T)} \\
& = 8T^2\mu^2\ln(2\cosh(\mu/2T)) +
16T^3 \mu[ \mbox{Li}_2[-e^{\mu/T}]- \mbox{Li}_2[-e^{-\mu/T}]] -
24T^4 [ \mbox{Li}_3[-e^{\mu/T}]+ \mbox{Li}_3[-e^{-\mu/T}]]
\end{split}
\end{equation}
and
\begin{equation}
\label{I}
\begin{split}
I& =\int \limits_{-\infty}^\infty\frac{d\omega(\omega-\mu)|\omega|}
{\cosh^2((\omega-\mu) /2T)} =
8T^3[ \mbox{Li}_2[-e^{-\mu/T}]- \mbox{Li}_2[-e^{\mu/T}]] -
8T^2\mu\ln(2\cosh(\mu/2T)) .
\end{split}
\end{equation}
These rather complicated expressions for  $I$ and $J$
can be expressed via elementary functions
in the limit $\mu \gg T$. We begin with the more simple case of $I$.
Using the identity \cite{Wolfram}
\begin{equation}
\label{Li2}
\mbox{Li}_2(z)=-\mbox{Li}_2(1/z)-\frac{1}{2}\ln^2(-z)-\frac{\pi^2}{6},
\quad z \notin (0,1),\qquad \mbox{Li}_2(0) = 0,
\end{equation}
we obtain that in the limit $\mu \gg T$
\begin{equation}
\mbox{Li}_2[-e^{-\mu/T}]- \mbox{Li}_2[-e^{\mu/T}] =
2 \mbox{Li}_2(-e^{-\mu/T})+ \frac{1}{2}\ln^2(e^{\mu/T})+ \frac{\pi^2}{6}
\simeq \frac{\mu^2}{2 T^2} + \frac{\pi^2}{6}.
\end{equation}
Thus the final expression for $I$ reads as
\begin{equation}
\label{I1}
I \simeq \frac{4 \pi^2}{3} T^3, \qquad \mu \gg T.
\end{equation}
To simplify $J$, in addition to Eq.~(\ref{Li2}),
we use the corresponding identity for $\mbox{Li}_3(z)$ \cite{Wolfram}:
\begin{equation}
\mbox{Li}_3(z)= \mbox{Li}_3(1/z)-\frac{1}{6}\ln^3(-z)-\frac{\pi^2}{6} \ln(-z),
\quad z \notin (0,1),
\qquad \mbox{Li}_3(0) = 0,
\end{equation}
and for $\mu \gg T$ we obtain that
\begin{equation}
\label{J1}
\begin{split}
J  \simeq   \frac{4 \pi^2}{3} T^3 \mu, \qquad \mu \gg T.
\end{split}
\end{equation}
Putting Eqs.~(\ref{J1}) and (\ref{I1}) along with
Eq.~(\ref{sigma.T>Gamma}) into
Eq.~(\ref{thermal.conductivity.general.final3}), we arrive at the
final result (\ref{kappa.T>Gamma}).


\end{document}